%% file: main.tex
\documentclass[twocolumn, tighten]{aastex62}
\usepackage{xspace}
\usepackage{graphicx}
\usepackage{amssymb}
\usepackage{amsmath}
\usepackage{subfigure}
\usepackage{hyperref}
\usepackage{footmisc}

\usepackage{natbib} 

\newcommand{\myemail}{zpace@astro.wisc.edu}

\newcommand{\mplv}{MPL-8\xspace}
\newcommand{\mplvngal}{6779\xspace}
\newcommand{\mplvfull}{MaNGA Product Launch 8\xspace}

\newcommand{\logmstar}{\ensuremath{\log M^* {\rm [M_{\odot}]}}}

\newcommand{\hi}{\ion{H}{1}\xspace}

\newcommand{\ha}{H\ensuremath{\alpha}\xspace}
\newcommand{\hb}{H\ensuremath{\beta}\xspace}
\newcommand{\hg}{H\ensuremath{\gamma}\xspace}
\newcommand{\hd}{H\ensuremath{\delta}\xspace}

\newcommand{\metlin}{\ensuremath{(O/H)^{\dagger}}\xspace}

\newcommand{\met}{\ensuremath{(O/H)^*}\xspace}
\newcommand{\metdec}{\ensuremath{\Delta \met}\xspace}
\newcommand{\metdisp}{\ensuremath{\tilde{\sigma} \met}\xspace}
\newcommand{\hifrac}{\ensuremath{\frac{M_{\rm HI}}{M^*}}\xspace}

\newcommand{\cnxs}{\ensuremath{{\rm CNXS}}\xspace}

\defcitealias{chen_pca}{C12}
\defcitealias{pettini_pagel_2004}{PP04}
\defcitealias{pilyugin_grebel_2016}{PG16}

\revised{\today}
\submitjournal{ApJL}

\shorttitle{Tests of a simple inflow model}
\shortauthors{Pace et al.}
\watermark{draft}

\begin{document}

\title{SDSS-IV/MaNGA: Can impulsive gaseous inflows explain steep oxygen abundance profiles \& anomalously-low-metallicity regions?}

\correspondingauthor{Zachary J. Pace}
\email{\myemail}

\author[0000-0003-4843-4185]{Zachary J. {Pace}}
\affil{Department of Astronomy, University of Wisconsin-Madison, 
       475 N Charter St., Madison, WI 53706}

\author[0000-0003-3097-5178]{Christy {Tremonti}}
\affil{Department of Astronomy, University of Wisconsin-Madison, 
       475 N Charter St., Madison, WI 53706}
       
\author{Adam L. {Schaefer}}
\affil{Department of Astronomy, University of Wisconsin-Madison, 
       475 N Charter St., Madison, WI 53706}

\author[0000-0002-3746-2853]{David V. {Stark}}
\affil{Department of Physics \& Astronomy, Haverford College, Marian E. Koshland Integrated Natural Science Center, 370 Lancaster Ave., Haverford, PA 19041}

\author[0000-0001-5457-6062]{Catherine A. Witherspoon}
\affil{Department of Astronomy, University of Wisconsin-Madison, 
       475 N Charter St., Madison, WI 53706}

\author[0000-0003-0846-9578]{Karen L. {Masters}}
\affil{Department of Physics \& Astronomy, Haverford College, Marian E. Koshland Integrated Natural Science Center, 370 Lancaster Ave., Haverford, PA 19041}

\author[0000-0002-7339-3170]{Niv {Drory}}
\affil{McDonald Observatory, The University of Texas at Austin, 1 University Station, Austin, TX 78712, USA}

\author[0000-0002-9808-3646]{Kai {Zhang}}
\affil{Lawrence Berkeley National Laboratory, 1 Cyclotron Road, Berkeley, CA 94720, USA}

\begin{abstract}
Gaseous inflows are necessary suppliers of galaxies' star-forming fuel, but are difficult to characterize at the survey scale. We use integral-field spectroscopic measurements of gas-phase metallicity and single-dish radio measurements of total atomic gas mass to estimate the magnitude and frequency of gaseous inflows incident on star-forming galaxies. We reveal a mutual correlation between steep oxygen abundance profiles between $0.25-1.5 ~ R_e$, increased variability of metallicity between $1.25-1.75 ~ R_e$, and elevated HI content at fixed total galaxy stellar mass. Employing a simple but intuitive inflow model, we find that galaxies with total stellar mass less than $10^{10.1} ~ {\rm M_{\odot}}$ have local oxygen abundance profiles consistent with reinvigoration by inflows. Approximately 10-25\% of low-mass galaxies possess signatures of recent accretion, with estimated typical enhancements of approximately 10-90\% in local gas mass surface density. Higher-mass galaxies have limited evidence for such inflows. The large diversity of HI mass implies that inflow-associated gas ought to reside far from the star-forming disk. We therefore propose that a combination of high HI mass, steep metallicity profile between $0.25-1.5 ~ R_e$, and wide metallicity distribution function between $1.25 - 1.75 ~ R_e$ be employed to target possible hosts of inflowing gas for high-resolution radio follow-up.
\end{abstract}



\input{intro.tex}

\input{data.tex}

\input{abund_meas_grad.tex}
\input{gradient_trends.tex}

\input{discussion.tex}

\input{acknowledgements.tex}

\software{Astropy \citep{astropy}, matplotlib \citep{matplotlib}, \texttt{scikit-learn} \citep{scikit-learn}, \texttt{cenken} \citep[\texttt{R} package: ][]{helsel_05_nondetects, akritas_murphy_lavalley_95}}

\bibliography{main}


\listofchanges

\end{document}

%% file: intro.tex
\section{Introduction}
\label{sec:intro}


Galaxies must exchange gas with their immediate environments: the continuous consumption of gaseous reservoirs through star formation implies a need for replenishment \citep{kennicutt_evans_2012_sf-review}. For example, to sustain the Milky Way's current star formation rate (SFR) of $\sim 1.5 {\rm M_{\odot} yr^{-1}}$ over the past 8 Gyr (a factor of several greater than the depletion time, the reciprocal of the star formation efficiency), gas capable of forming stars should be introduced at a similar rate \citep{snaith_2015_mw-sfr, licquia_newman_2015_mw}. This is also true near the peak of the cosmic star formation history, $z=0-2$ \citep{tacconi_2013}, indicating that refueling is important across cosmic time. Furthermore, the empirical relationship between galaxies' star formation and gas content implies that the resupplying process is integral to galaxies' evolution through time \citep{kennicutt_1998}. This need for additional fuel is exacerbated by star-formation-driven outflows: winds from massive stars and supernovae are capable of launching outflows into galaxy haloes at rates greater than the SFR itself \citep{heckman_1990_outflows, rubin_2014_outflows, chisholm_2018}, further depleting available gas reservoirs. Finally, cosmological simulations seem to demand inflows in order to reproduce galaxies' SFRs and buildup of stellar mass over cosmic time: the inflowing gas present in simulations can reach the disks of galaxies with $M^* < 10^{10.3} {\rm M_{\odot}}$ without being shock-heated, implying that it will be detectible as HI \citep{keres_2005_cold-accretion, dekel_birnboim_06_cold-accretion}.

While outflows from galaxy disks are well-studied phenomena in the extragalactic environment, inflows have proven more elusive to direct detection at the survey scale. High-column-density, inflowing structures such as high-velocity clouds (HVCs--\citealt{wakker_2004_hvc_bookchapter}) are ubiquitous in the Milky Way's immediate vicinity, and similar structures ought to exist in other haloes, as well. However, direct detection of HI in extragalactic HVC analogs is not practical, since the radio array configurations with sufficient spatial resolution lack the sensitivity to detect low-column-density structures. Furthermore, it is difficult to establish an unassailable link between gas in the vicinity of galaxies and the star formation in the galaxy itself: after all, cold gas is pervasive in galaxy haloes regardless of level of star formation activity \citep[e.g.,][]{bieging_78_etg-HI, sanders_80_etg-HI, zhang_2019_quiescent-HI}. Projection effects, unknown ionization state, obscuration on the far side of a dusty disk, and the paucity of metals all contribute to challenges in observing potential gaseous inflows. Targeted studies of individual galaxies have produced interesting examples of low-metallicity star formation indicative of ongoing accretion \citep{howk_2018_ngc4013-1, howk_2018_ngc4013-2, sanchez-almeida_2014}; but larger samples across a wider range of galaxy properties, conditions, and environments are needed. The difficult aspects of this open question must be addressed: to quantify the impact of inflowing gas is to ground one branch of the baryon cycle.

Since direct measurements of inflow are presently impractical at the survey scale, we must seek out its secondary effects, specifically on galaxies' chemistry: a galaxy's gas content is tied to the abundance of heavy elements in its interstellar medium (ISM), and therefore also to the buildup of stellar populations over cosmic time. Models of galaxy chemical evolution see gas mass fraction, gas-phase metallicity\footnote{Throughout this work, we use the terms ``oxygen abundance", ``gas-phase metallicity", and simply ``metallicity" interchangeably.}, and stellar mass as manifestations of galaxies' active transformation of gas into stars, with heavier elements released back into the ISM during the final stages of massive stars' lives \citep{tinsley_72, tinsley_73, vincenzo_2016_yields}. The observational evidence for this interplay is abundant in statistical samples of star-forming galaxies, with central metallicity increasing with stellar-mass \citep{tremonti_mz}, and gas fraction decreasing as stars and metals accumulate \citep{hughes_2013-gfrac-met}. It appears necessary to have some combination of gaseous inflows and outflows to explain the chemical abundances of old stars in the Milky Way \citep{spitoni_2019_mw-chemistry} and gas-phase abundances in other galaxies \citep{lilly_2013_gasreg}.

Because inflows may be brief and localized, rather than smooth in space and time, they may produce detectable chemical signatures in the disk. In general, a star-forming galaxy's metallicity decreases as galactocentric radius increases \citep{oey_kennicutt_1993, zaritsky_1994, sanchez_2014_metgrad, belfiore_2017_manga-metgrad, poetrodjojo_18, sanchez_19_review}. Reports differ regarding whether the slope of a galaxy's oxygen abundance profile (the metallicity gradient) correlates with its total stellar mass \citep{sanchez_2014_metgrad, belfiore_2017_manga-metgrad, zinchenko_2019, mingozzi_2020}. It has been argued that characteristic metallicity gradients emerge from inside-out galaxy formation \citep{prantzos_boissier_2000_metgrad}; but, they may simply emerge from the gaseous reservoirs' evolution \emph{at a local scale} \citep{moran_2012, zhu_2017_leakybox_smsd, barrera-ballesteros_2018, sanchez_almeida_localfmr, bluck_2019_global-local}. 

While common and statistically well-characterized on average, metallicity gradients are not perfectly uniform: chemical abundance does appear to vary azimuthally as well as radially, despite the rapid timescale (within one galactic rotation period) on which metals ejected from massive stars are thought to mix with the surrounding ISM \citep{petit_2015_azimixing}. In the Milky Way, for instance, abundance gradients measured at different azimuth angles have been found to differ by $\pm 0.03 ~ {\rm dex ~ kpc^{-1}}$ \citep{balser_2015_MWmetgrad}; but it is not clear whether this behavior extends to all star-forming galaxies as a population \citep[for a diversity of views, see][]{kreckel_2016_interarm, sanchez-menguiano_2016, vogt_2017_metgrad-azi, ho_2017_metgrad-azi, ho_2018_metgrad-azi}. Recently, \citet{kreckel_2019_phangs_metgrads} have reported typical metallicity dispersions of $0.03 - 0.05 ~ {\rm dex}$ at fixed radius in a sample of galaxies observed with VLT/MUSE. Furthermore, deviations from single gradients in individual galaxies have been observed---albeit in somewhat smaller samples---with breaks in the radial metallicity profiles separating the disks' innermost regions, intermediate radii, and outskirts \citep{sanchez-menguiano_2019}. 

Simultaneously, integral-field surveys such as MaNGA have yielded reports of anomalously-low-metallicity (ALM) regions, atypically-metal-poor areas at $\sim {\rm kpc}$ scales: \citet{hwang_2019_manga_almrs} finds a sizable ALM sample (defined as having oxygen abundance at least $0.14 ~ {\rm dex}$ below the mean metallicity for all MaNGA spaxels at the same stellar mass surface density and total galaxy stellar mass) in the MaNGA sample. About 25\% of local star-forming galaxies reportedly exhibit these characteristics, preferring galaxies below $10^{10} {\rm M_{\odot}}$ and morphologically-disturbed galaxies; and about 10\% of all MaNGA star-forming spaxels have the ``ALM" designation. The suggested explanation is the rapid, impulsive (``bursty") accretion of gas from the halo which fuels star formation.

Despite the recent evidence that has emerged for localized inflows onto star-forming disks, the details remain indistinct. Cosmic filaments, the source for inflowing gas in cosmological simulations, are many times the size of galaxies' star-forming disks \citep{martin_inflows}; and some questions remain about whether their associated inflows could effect a sustained and detectible depression of metallicity. There are now efforts to replicate survey-scale measurements of metallicity gradients in synthetic data from cosmological simulations \citep{hemler_2020_tng50-metgrad}, undertakings which will be enhanced with improved forward-modelling of metallicity indicators and GMC-scale physics. Ultimately, though, the duty cycle of inflows (i.e., the fraction of time that the average gas reservoir actively accretes gas) is also relatively ill-constrained; and thus, to establish the definitive link between ALM gas and an actual inflow, the local gas reservoir should be characterized---an important deficiency in the current generation of observations. At present, the best indication of local gas supplies at the survey scale is global (single-dish) HI measurements. In addition, given the relatively short mixing timescale of heavy elements in the ISM, the width of the metallicity distribution function at fixed radius (or within a narrow radial range) could be employed as a tracer of newly-introduced gas with non-ambient metallicity. This work will address such opportunities.

This study seeks to link the diversity of radial metallicity profiles \& the width of the metallicity distribution function at fixed radius with galaxies' atomic gas reservoirs, by exploring the mutual correlations between radial oxygen abundance slope, oxygen abundance dispersion, total HI mass, and total stellar mass; and employs a conceptual model of local dilution to estimate plausible enhancements to star-forming gas reservoirs at the local (kpc) scale. In Section \ref{sec:data}, we describe the MaNGA resolved spectroscopy and three related value-added catalogs (VACs) which provide measurements of total galaxy stellar-mass, measurements of disk effective radius, and total HI mass. In Section \ref{sec:abund_meas_grad_dec}, we describe the strong-line metallicity calibration used; measure radial metallicity profiles with a metallicity decrement (a replacement for the more traditional gradient); measure azimuthal metallicity variations using the width of the metallicity distribution in a narrow annulus; and delineate the sample selection. In Section \ref{sec:trends}, we report the trends between HI mass, metallicity decrement, metallicity distribution width, and total stellar mass. In Section \ref{sec:modeling}, we describe the local dilution of star-forming gas reservoirs with an intuitive model. In Section \ref{sec:disc}, we summarize our results \& their implications. Throughout this work, we adopt the nine-year WMAP cosmology \citep{wmap9}, and a \citet{kroupa_imf_01} stellar initial mass function (IMF).

%% file: data.tex
\section{Data}
\label{sec:data}

This work uses integral-field spectroscopic (IFS) data from the MaNGA survey \citep{bundy15_manga}, part of SDSS-IV \citep{blanton_17_sdss-iv}. MaNGA will observe more than 10,000 nearby galaxies ($0.01 < z < 0.15$) from the NASA-Sloan Atlas \citep[NSA, ][]{blanton_11_nsa}, with an approximately-uniform distribution in $i$-band absolute magnitude, resulting in a roughly-flat distribution in $\log M^*$, and is approximately volume-limited within a given redshift range \citep{manga_sample_wake_17}. Two-thirds of observed galaxies are drawn from the ``Primary+" sample (coverage to at least 1.5 $R_e$); and the remainder come from the ``Secondary" sample (covered to at least 2.5 $R_e$). The spectroscopic data used in this study come from \mplvfull (\mplv), an internally-released dataset nearly identical to SDSS Data Release 16 \citep{sdss_dr16}, and containing \mplvngal galaxies.

MaNGA spectroscopic observations cover the wavelength range of 3600 to 10300 $\mbox{\AA}$ with $d\log\lambda \sim 10^{-4}$ ($R \sim 2000$), and use the BOSS spectrograph \citep{smee_boss_instrument, sdss_boss_dawson_13}, an instrument at the SDSS 2.5-meter telescope at Apache Point Observatory \citep{gunn_sdss_telescope}. To achieve uniform spatial sampling, the spectrograph is coupled to closely-packed fiber hexabundles, each containing between 19 and 127 fibers \citep{manga_inst}. Sky-subtraction and spectrophotometric calibration are accomplished using single fibers and smaller hexabundles \citep{manga_drp, manga_spectrophot}. All hexabundles and sky fibers are inserted into a plugplate affixed to the focal plane \citep{sdss_summary}. Sets of three ``dithered" pointings compose the MaNGA observations, and to form the datacubes (\texttt{CUBE} products), these observations are rectified to a spatial grid in the plane of the sky, with spaxel size 0.5" by 0.5" and seeing-induced PSF having an $i$-band FWHM $\sim 2.5$" \citep{manga_obs, manga_progress_yan_16, manga_drp}. The MaNGA Data Analysis Pipeline \citep[DAP, ][]{manga_dap} measures stellar kinematics, emission-line fluxes \citep{belfiore_2019_mangadap-emissionline}, and stellar spectral indices for individual spaxels.

This work builds on the results of two MaNGA Value-Added Catalogs (VACS). First, we use estimates of total galaxy stellar-masses from the MaNGA-PCA project, which used an orthogonal spectral basis set trained on realistic SFHs to obtain robust resolved galaxy stellar masses \citep{pace_19a_pca}; the resolved masses were then aperture-corrected to form a catalog of total stellar-masses \citep{pace_19b_pca}. These stellar masses are likely much more reliable than those included in the MaNGA targeting catalog, since galaxy-averaged light is more apt to ``miss" stellar-mass in dusty environments and other low-flux regions \citep{zibetti_2009, sorba_sawicki_15, pace_19b_pca}. Second, the MaNGA PyMorph DR15 photometric catalog provides parameters obtained from S\'{e}rsic and S\'{e}rsic-Exponential fits to MaNGA galaxies' plane-of-sky surface-brightness profiles \citep{fischer_2019_pymorph}. This allows radial abundance trends to be computed with respect to the disk (a more fundamental unit of chemical evolution), rather than the disk plus the bulge.

Finally, we include single-dish atomic hydrogen (HI) mass measurements and upper-limits: the GASS \citep{catinella_2010_GASS} and ALFALFA \citep{haynes_2018_alfalfa} surveys form the small archival portion of the HI measurements. The majority of measurements come from the HI-MaNGA project \citep{masters_19_himanga, goddy_2020_gbtcal, stark_himanga}, an observational campaign carried out with the Green Bank Telescope. This program targets MaNGA galaxies at $cz < 15,000 ~ {\rm km s^{-1}}$, but regardless of their morphology, with an intended depth of 1.5 mJy at $10 ~ {\rm km s^{-1}}$ (after spectral smoothing), resulting in a stellar-mass distribution of targets peaking at $M^* \sim 10^{9.8} {\rm M_{\odot}}$ \citep[see Figure 1 of][]{masters_19_himanga}. GBT HI observations were translated into HI mass estimates, and in the case of non-detections, mass upper-limits were estimated using the observational noise and assuming a rotation-curve with of $200 {\rm km ~ s^{-1}}$. In total, 3413 MaNGA galaxies have measured HI masses or upper-limits.

%% file: abund_meas_grad.tex
\section{Abundance calibrations and radial decrements}
\label{sec:abund_meas_grad_dec}

We measure gas-phase metallicities of individual MaNGA spaxels using the ratios between the fluxes of strong nebular emission lines, since the ``direct" ($T_e$) method requires much deeper spectroscopy at the high expected metallicities of MaNGA spaxels. We rely on one of the strong-line calibrations of \citet[][hereafter, \citetalias{pilyugin_grebel_2016}]{pilyugin_grebel_2016}, which matches several strong-line ratios to $T_e$ abundances over a reference sample of 313 HII regions. Specifically, we use the \texttt{R2} calibration, referred to hereafter as \texttt{PG16-R2}. The three strong-line ratios used to define the calibration in \citetalias{pilyugin_grebel_2016} are defined as follows:

\begin{itemize}
    \item $R_2 = \frac{\rm F([OII]_{3727}) + F([OII]_{3729})}{\rm F(H \beta)}$
    \item $R_3 = \frac{\rm F([OIII]_{4959}) + F([OIII]_{5007})}{\rm F(H \beta)}$
    \item $N_2 = \frac{\rm F([NII]_{6548}) + F([NII]_{6584})}{\rm F(H \beta)}$
\end{itemize}

Generally, the strong-line ratios $R_2$ \& $R_3$ (as well as their sum, generally notated $R_{23}$), are double-valued; that is, a single value can indicate one of two values of oxygen abundance. In the \texttt{PG16-R2} calibration, this degeneracy between ``upper" and ``lower" branches is broken using the $N_2$ ratio (previous work has cautioned against using nitrogen-based line ratios to determine oxygen abundance, because the abundance ratio of N/O is variable, even at fixed oxygen abundance---but adopting it for the coarse task of deciding the $R_{23}$ branch is safer). The accuracy of the resulting oxygen abundance estimate is $\lesssim 0.1~{\rm dex}$ over the range $7.0 \le 12 + \log{\frac{O}{H}} \le 9.0$, and \citet{schaefer_19_ohno} has found the \texttt{PG16-R2} calibration to be less susceptible to N/O variations than other strong-line calibrations. Since variations in excitation (ionization parameter) can impact reliability of abundance estimates, this calibration adjusts for this effect using the ratio between $R_2$ \& $R_3$. \citep{pilyugin_grebel_2018_manga} concluded that for MaNGA spectra, the \texttt{PG16-R2} calibration does not suffer from the same excitation-dependent deficiencies that purely $N_2$-dependent calibrations do.

The metallicity for a particular spaxel is estimated according to a Monte Carlo randomization of attenuation-corrected emission-lines. The first four Balmer emission lines (\ha, \hb, \hg, \hd) are used to estimate the attenuation law, assuming a \citet{charlot_fall_00} two-component dust model. The best-fitting dust parameters $\tau_V$ and $\mu$ (along with their covariances) are found through Levenberg-Marquardt optimization. The line-specific attenuation correction-factors are then randomized according to the covariance matrix of the best-fit parameters, with 1000 draws total. The observed emission-line fluxes are likewise independently resampled 1000 times according to their reported uncertainties, and multiplied with the resampled dust-corrections. The combination of these draws for all emission lines is used to approximate the distribution of possible strong-line ratios $R_2$, $R_3$, and $N_2$, which themselves are used to approximate draws from the distribution of oxygen abundance. This process is repeated for each spaxel in the field of view. The median of those 1000 draws from the oxygen abundance distribution is taken as the fiducial oxygen abundance for a spaxel, \met (notation employed for consistency with \citealt{pilyugin_grebel_2016}); with the median absolute deviation of that distribution is taken as a measure of the dispersion. Both are are logarithmic quantities.

\subsection{Radial decrement definition}
\label{subsec:bintervals}

Since the star-forming disk is the main engine driving galaxy chemical evolution, the most useful radial unit is a \emph{disk effective radius} ($R_e^d$) rather than a \emph{total effective radius} ($R_e^t$). We use the bulge-disk decompositions from the MaNGA PyMorph photometric catalog \citep{fischer_2019_pymorph} to assign a radial coordinate to all spaxels based on the measured disk effective radius (hereafter, the shorthand $R_e$ refers to a disk effective radius, unless otherwise specified). In practice, radial oxygen abundance trends fitted to ensembles of individual spaxels tend to be dictated strongly by the measurements in the outermost $\sim 0.5 R_e$, since the number of spaxels within a radial interval rises in proportion to the distance from the center of the galaxy. For the case of an abundance gradient with non-constant slope, the central abundance would also be improperly estimated. In \citet{belfiore_2017_manga-metgrad}, metallicity measurements were binned in intervals of 0.1 $R_e$ and an unweighted least-squares fit made, thereby standardizing the contribution at various radii to the gradient fit. Instead, we define a radial metallicity ``decrement" between two widely-spaced annuli, as $\Delta_{in}^{out} = \met_{in} - \met_{out}$; in other words, a positive decrement indicates that oxygen abundance decreases radially (unlike a positive metallicity gradient, which conventionally signifies a radially-increasing metallicity profile).

The bounds for the radial bins are chosen to minimize the adverse effects of blurring by the PSF. We exclude spaxels along or close to the minor axes of inclined galaxies, where the PSF-induced smoothing in the data will ``smear out" radial variations most severely: galaxies with observed minor-to-major axis ratios under 0.33 are excluded entirely, since dust-correction becomes extremely problematic at inclinations greater than 75 deg \citep{pellegrini_2019_warpfield}. At axis ratios greater than 0.75, it is difficult to accurately determine a galaxy's axis ratio, and so all spaxels are included. In the intermediate range ($0.33 < b/a < 0.75$), the azimuthal angle from the major axis determines the acceptance or rejection of a spaxel, with the angle rising linearly from 20 degrees at $b/a = 0.33$ to 90 degrees (i.e., no restriction) at $b/a = 0.75$.

\begin{figure*}
    \centering
    \includegraphics[width=5 in]{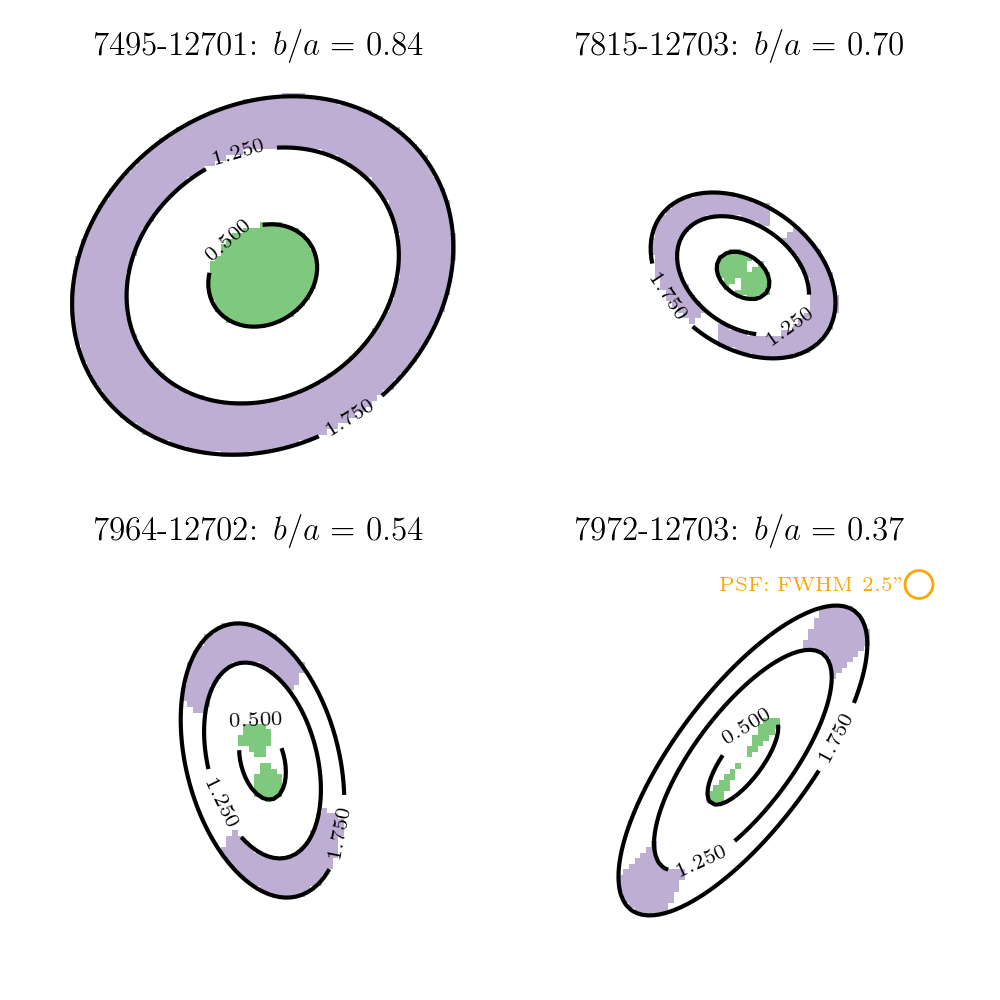}
    \caption{The radial binning scheme described above, applied to four MaNGA galaxies with decreasing disk aspect-ratios. Pixels are colored according to their use in radial bins: light green indicates radial bin 0 ($0.0 - 0.5~R_e$), and lavender indicates radial bin 1 ($1.25-1.75~R_e$). Pixels colored white indicate locations not included in any bin, due to either radial coordinate or azimuthal angle away from the major axis. These restrictions are applied in addition to those on data-quality (see Section \ref{subsubsec:data_quality}). Contours are in units of $R_e$, and are labeled at the outer boundary of bin 0, as well as the inner \& outer boundaries of bin 1. The spatial resolution element (FWHM) is visualized as an orange circle in the bottom-right panel: as inclination increases, the PSF samples an increasing diversity of galaxy radii along the minor axis.}
    \label{fig:radial_bins}
\end{figure*}

Though we might desire to maximize the number of radial bins available, there is limited utility to reducing bin width smaller than the PSF size. Furthermore, a larger radial interval allows more spaxels to be aggregated. Most important for our purposes is the observation that both the MaNGA Primary and Secondary samples are relatively clean from PSF-induced contamination at radial separations greater than $0.5~R_e$. We elect to use radial bins with widths of $0.5~R_e$, and spaced apart by $0.75~R_e$ (see Figure \ref{fig:radial_bins}). In this work, we consider bin limits of [0.0, 0.5] $R_e$ (radial bin 0) and [1.25, 1.75] $R_e$ (radial bin 1).

\subsection{Sample \& Data quality}
\label{subsubsec:data_quality}

Our base sample of galaxies is the PyMorph VAC (4266 galaxies), almost all of which have total stellar masses from the MaNGA-PCA VAC. Of these, 1152 (2080) have \hi mass estimates (upper-limits) from either targeted GBT follow-up \citep{masters_19_himanga, goddy_2020_gbtcal}, ALFALFA \citep{haynes_2018_alfalfa}, or GASS/GASS-low \citep{catinella_2010_GASS}, accessed in that descending order of priority. 

We adopt the following constraints on individual spaxels within galaxies, all of which must be met in order for the spaxel to be included in the radial fits and decrement calculation. These cuts are applied in addition to the radial and azimuthal restrictions described above.

\begin{itemize}
    \item Signal-to-noise cuts: We require $S/N(\ha) > 5$, $S/N(\hb) > 3$, $S/N([{\rm O ~ III}]) > 3$, and $S/N([{\rm N ~ II}]) > 3$, since these lines form the basis for the selection of star-formation-dominated spaxels.
    \item Excitation cuts: We requires spaxels to lie in the star-formation-dominated portion of the $[{\rm N ~ II}] / \ha$-versus-$[{\rm O ~ III}] / \hb$ excitation diagram, i.e., below the \citet{kauffmann_03_agn} and \citet{kewley_dopita_01} demarcation lines.
    \item Diffuse ionized gas (DIG) rejection: \citet{lacerda_2018_califa_dig} finds that when EW(\ha) is less than 3$\mbox{\AA}$, DIG dominates the emission spectrum. We select only spaxels with $EW(\ha) \ge 3 {\rm \AA}$.
    \item Restrictions on resampled oxygen abundances: The Monte Carlo-resampled distribution of the oxygen abundance (obtained from the line-ratio calibration) must have a median absolute deviation (MAD) $< 0.3 ~ {\rm dex}$, and must lie in the range [7.0, 9.0] (where the \texttt{PG16-R2} calibration is well-characterized).
\end{itemize}

In order for an individual galaxy to be included for the purposes of radial metallicity trends, there must be at least 8 (10) unmasked spaxels in radial bin 0 (radial bin 1). With these thresholds set, we are left with 252 (110) galaxies having \hi masses (upper-limits). A further 273 galaxies passing all spectroscopic data-quality criteria above (bullet-points), but which were not targeted in the HI follow-up, are included separately.

\subsection{Decrements versus previously-measured gradients}

Here we consider how metallicity decrements between radial annuli compare to previous measurements of abundance gradients. The relationship between a radial decrement and a gradient is intuitive: after adopting fiducial endpoints of $0.25 ~ R_e$ and $1.5 ~ R_e$ (the midpoints of the inner and outer radial intervals) and negating to reflect the sign difference (a positive decrement and a negative gradient each reflect a decreasing radial metallicity profile), the median measured metallicity gradients obtained in \citet[a recent study using MaNGA data]{mingozzi_2020} can be transformed into decrements, for comparison to this work. For this comparison only, the directly-measured gradients are multiplied by $R_e^d / R_e^t$, which approximately accounts for the differing slopes measured in ``disk" and ``total" radial coordinates. In addition, we fit (and similarly correct) all galaxies' individual radial metallicity profiles using an identical linear model to \citet{mingozzi_2020}, and using identical spaxel cuts to those employed for the decrements. In general, these two radial trends are very similar, with the exception of galaxies with steep gradients ($\nabla < -0.1 ~ {\rm dex ~ R_e^{-1}}$), for which the corresponding decrements are substantially smaller.

\begin{figure*}
    \centering
    \includegraphics{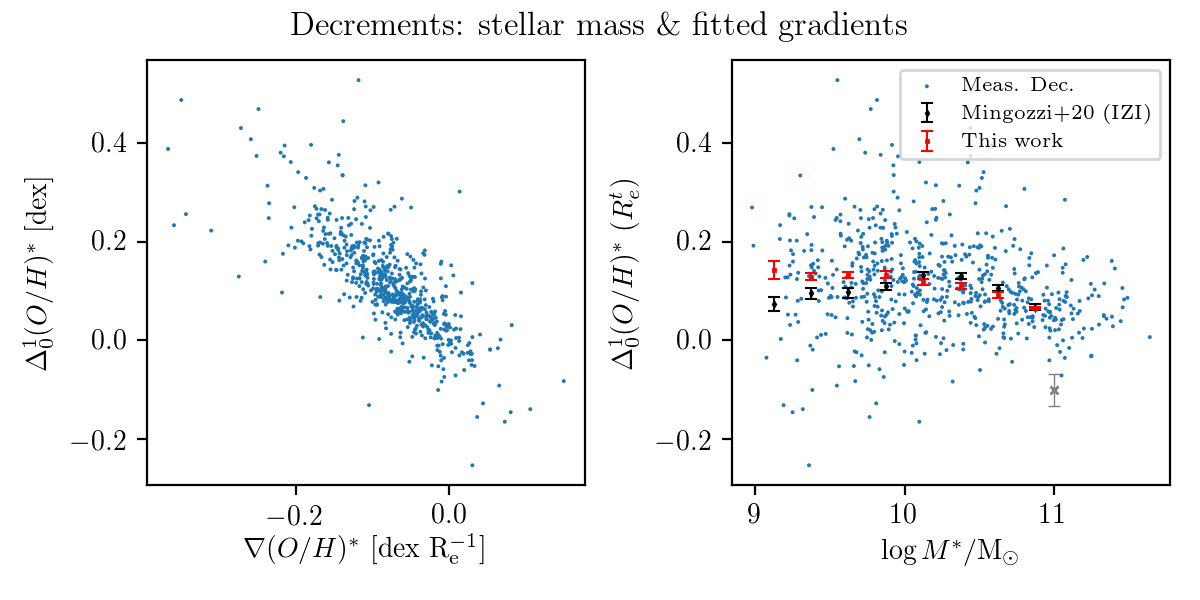}
    \caption{A summary of the comparison between mass-aggregated radial metallicity trends from \citet{mingozzi_2020}, and this work's radial metallicity decrements. \textbf{Left panel:} a comparison between the decrement obtained by direct computation using spaxel metallicity values, and a gradient performed by fitting the radial profile with a linear model (both in $R_d$ coordinates). The two measurements track relatively consistently at for shallow metallicity profiles ($\nabla > -0.1 ~ {\rm dex ~ R_d^{-1}}$); galaxies with steep profiles tend to have the gradient-converted metallicity decrement overestimate the directly-computed decrement. \textbf{Right panel:} a stellar-mass-dependent comparison of this work's metallicity decrements with those obtained by converting the averaged, population-level gradients of \citep{mingozzi_2020} into decrements; and converting this work's decrements into $R_e^t$ coordinates (also present in Table \ref{tab:decrement_mstar}). The directly-measured; typical decrement uncertainty is shown in gray; the average decrements from \citep{mingozzi_2020} are converted to decrements and shown as black points, with errorbars signifying the uncertainty on the mean gradient at that stellar mass; and the average decrements from this work are shown in red (errorbars signify uncertainty on the median decrement). \emph{Note: The adaptation of the \citet{mingozzi_2020}} gradients includes a sign change, since this work defines the metallicity decrement of a radially-decreasing metallicity profile as positive; whereas the definition of a gradient is typically the opposite.)}
    \label{fig:decfromgrad-dec-Re}
\end{figure*}

\begin{table}[]
    \centering
    \begin{tabular}{c|c}
        Bin Limits & Med. \metdec \\ \hline \hline
        $\logmstar < 9.25$ & $0.143 \pm$ 0.018 \\ \hline
        $9.25 \le \logmstar < 9.5$ & $0.130 \pm 0.008$ \\ \hline
        $9.5 \le \logmstar < 9.75$ & $0.133 \pm 0.007$ \\ \hline
        $9.75 \le \logmstar < 10.0$ & $0.134 \pm 0.006$ \\ \hline
        $10.0 \le \logmstar < 10.25$ & $0.118 \pm 0.006$ \\ \hline
        $10.25 \le \logmstar < 10.5$ & $0.110 \pm 0.006$ \\ \hline
        $10.5 \le \logmstar < 10.75$ & $0.0925 \pm 0.0059$ \\ \hline
        $\logmstar > 10.75$ & $0.0657 \pm 0.0028$ \\ \hline
    \end{tabular}
    \caption{Median radial metallicity decrements in the mass bins adopted by \citet{mingozzi_2020}, as shown in Figure \ref{fig:decfromgrad-dec-Re}, in $R_e^t$ coordinates.}
    \label{tab:decrement_mstar}
\end{table}

Figure \ref{fig:decfromgrad-dec-Re} confronts the aggregate metallicity profile trends found in \citep{mingozzi_2020}, with galaxy-by-galaxy decrements found in this work (both measured directly then converted to $R_e^t$ coordinates, and obtained by conversion from gradients measured in previous work). Table \ref{tab:decrement_mstar} likewise reflects the directly-measured decrements from this work, in \emph{total effective radius} ($R_e^t$) coordinates. In making a comparison to previous work, it is crucial to note that the \citet{mingozzi_2020} points and bounds represent constraints on the \emph{median radial abundance trend} at fixed mass, rather than the \emph{true distribution of metallicity gradients}. Across the entire mass range, the \citet{mingozzi_2020} gradient-derived decrements lie well within the distribution of individual galaxies' directly-measured decrements. Though the medians correspond acceptably at high mass ($\logmstar \gtrsim 10$), the two classes of measurements appear less compatible at low stellar masses. Furthermore, this match is not substantially improved by using converting the gradients found above into decrements. Some combination of the following effects are likely responsible for such discrepancies:
\begin{itemize}
    \item The converted gradients from \citet{mingozzi_2020} rely on a Bayesian framework to determine each spaxel's metallicity \citep[\texttt{IZI:}][]{blanc_2015_izi}. The full effects of calibration-related systematics have not been fully isolated \citep{belfiore_2017_manga-metgrad}, but \citet{yates_2020_lgalaxies-sam-metgrad} notes that low-mass galaxies are more susceptible to indicator-induced metallicity gradient uncertainties.
    \item \citet{mingozzi_2020} fit the radial gradients only between $0.5-2.0 ~ R_e^t$, excluding nearly the full breadth of the innermost radial annulus adopted in this work, in order to sidestep non-linear radial profiles (in particular, the well-known flattening of the abundance profile inward of $1.0 ~ R_e$---see \citealt{mast_2014_califa-ifs-resolution} and \citealt{sanchez_19_review}, Figure 15 \& \citet{mingozzi_2020}, Figure 12). This is an unlikely source for systematics at low mass, since such flattening primarily occurs in more massive galaxies.
    \item The radial abundance trends are measured at the sample level in \citet{mingozzi_2020}. To wit, the \citet{mingozzi_2020} median gradients are found by first aggregating the spaxels of all galaxies within a mass interval into radial bins, and then fitting one linear regression against galactocentric radius for each mass interval. This pre-fit aggregation could conceivably reflect different behavior than the underlying distribution of galaxies possesses.
    \item Low-mass galaxies have gradient measurements somewhat susceptible to changes in metallicity indicators; and low-mass galaxies' photometric \& structural irregularities also may introduce problems characterizing radial abundance profiles. In aggregate, though, this tends to draw fitted gradients towards zero \citep{yates_2020_lgalaxies-sam-metgrad}, and could act in concert with the aggregate fitting described in the previous bullet-point to draw the \citet{mingozzi_2020} gradients away from the true population median. Interestingly, this study finds non-flat metallicity profiles for low-mass galaxies \emph{despite no explicit cut being made on morphology \& structure (unlike \citet{yates_2020_lgalaxies-sam-metgrad}}.
\end{itemize}

Ignoring for the moment the breadth of metallicity decrements present at fixed stellar mass, decrements' aggregate medians seem to reflect a slight flattening of typical metallicity profile slope with increasing stellar mass, a somewhat atypical result. \citep{yates_2020_lgalaxies-sam-metgrad} in particular find this effect still at play at $\logmstar \sim 9.25$, but a larger breadth of studies not this at $\logmstar > 10$, the precise interval where this work's decrements agree with those of \citet{mingozzi_2020} (but an interval which in this work is responsible for only about half of the total flattening observed). Understanding the low-stellar-mass regime and the systematics at play may provide a foothold into reconciling disagreements about the nature of abundance gradients at the survey scale.

In addition to the observed gradients from other integral-field studies, it is worth calling attention to ongoing efforts towards reproduce large-scale galaxy population behaviors in cosmological simulations. With an eye towards eventual, direct comparisons to observation, \citet{hemler_2020_tng50-metgrad} have assembled several sets of metallicity gradients measured from the IllustrisTNG cosmological simulation suite \citep{pillepich_2018-tngsim, weinberger_2017_tng-sim-feedback}. Though the simulation-based gradients are not synthetically observed in a manner compatible with current spectroscopic techniques; and though \citep{hemler_2020_tng50-metgrad} reports gradients in different radial coordinates; there are some useful points of comparison. For instance, like shown in Figure \ref{fig:decfromgrad-dec-Re}, simulated galaxies at $z \sim 0$ show a greater diversity of metallicity gradients at low stellar mas than at high stellar mass. As simulation suites grow in the future and synthetic observations follow, more direct comparisons to observed gradients at the survey scale may become practical.

%% file: gradient_trends.tex
\section{Abundance decrement trends}
\label{sec:trends}

Star-forming galaxies have been shown to possess similar metallicity gradients across a range of total stellar masses \citep{sanchez_2014_metgrad, belfiore_2017_manga-metgrad, mingozzi_2020}, with some evidence that gradients steepen with increasing galaxy stellar mass \citep{belfiore_2017_manga-metgrad, poetrodjojo_18, mingozzi_2020}. This work seeks to uncover the connection between a galaxy's radial metallicity profile and its global gas content. 

\subsection{Radial metallicity trends and HI}

\begin{figure*}
    \centering
    \includegraphics[width=\textwidth]{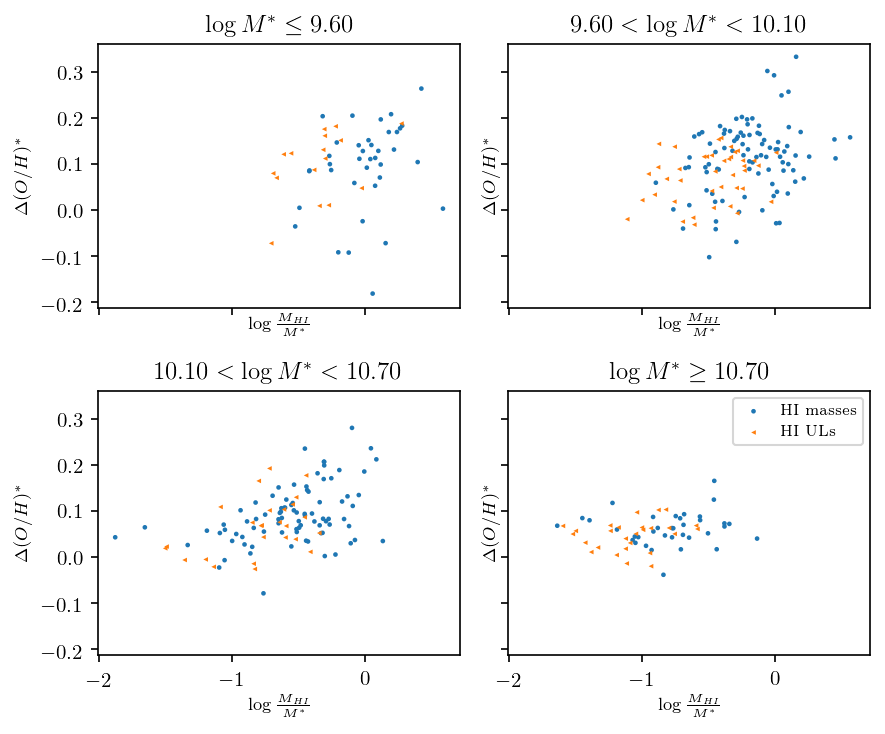}
    \caption{The radial oxygen abundance decrement between $0.0-0.5 ~ R_e$ \& $1.25-1.75 ~ R_e$ (\metdec) is shown on the abscissa, with respect to the HI mass normalized by the stellar mass ($\log M_{HI} / M^*$). Positive HI detections are shown as blue squares, and upper-limits as orange triangles: at fixed total stellar mass, the atomic gas mass upper-limits probe a lower gas-fraction branch of the galaxy population. There is a moderate correlation between HI mass fraction and oxygen abundance decrement strength at fixed galaxy total stellar mass, present most prominently at $M^* < 10^{10.7} {\rm M_{\odot}}$.}
    \label{fig:hifrac_dec01_subplotmstar}
\end{figure*}

Does HI content exert an influence over \metdec beyond its correlation with stellar mass? Figure \ref{fig:hifrac_dec01_subplotmstar} shows evidence of a positive correlation between HI mass fraction and radial gas-phase metallicity decrement, when galaxies are separated by their total stellar mass into four bins. The bins are set as the 16$^{th}$, 50$^{th}$, and 84$^{th}$ percentiles of the stellar-mass distribution of galaxies with either an H\textsc{i} mass measurement or upper-limit: $10^{9.6}$, $10^{10.1}$, and $10^{10.7}~{\rm M_{\odot}}$. The correlation manifests for galaxies with measured HI masses as well as upper-limits.

Since a steep decrement could emerge from a high central value, rather than a low outer value, a positive correlation between central metallicity and HI mass might feasibly bring about the observed correlation between strong decrement and HI mass fraction. In practice, though, galaxies with the largest HI fractions at fixed stellar mass also have \emph{lower-than-typical gas-phase metallicities} in their inner 0.5 $R_e$. This is consistent with the standard picture of galaxy chemical evolution \citep{tinsley_72, tinsley_73}, and qualitatively similar to other observed relations between gas fraction and oxygen abundance \citep{hughes_2013-gfrac-met}. Thus, we conclude that the correlation is not an emergent result of differing galaxy-to-galaxy abundance zeropoints.

The correlation we observe between a steep radial metallicity profile and a large HI mass is reminiscent of a result noted in a sample of DustPedia galaxies with archival MUSE integral-field spectroscopic observations \citep{devis_2019_dustpedia}. The DustPedia parent sample has a great breadth of photometric measurements (25 bands from UV to submillimeter); but there are only 75 galaxies with integral-field spectroscopic observations (from the MUSE archive), albeit with finer spatial sampling than MaNGA reaches. Radial metallicity trends are measured in \citet{devis_2019_dustpedia} with a linear fit, and are a true gradient over the entire galaxy, though with respect to $r_{25}$ (the radius of the $m_B = 25$ isophote). Though converting between gradients in the units of $r_{25}$ (DustPedia) and the decrements in the units of disk effective radii (this study) is not straightforward, we note with interest that there is some basis for a \metdec-HI correlation in the literature.

\begin{figure*}
    \centering
    \includegraphics[width=\textwidth]{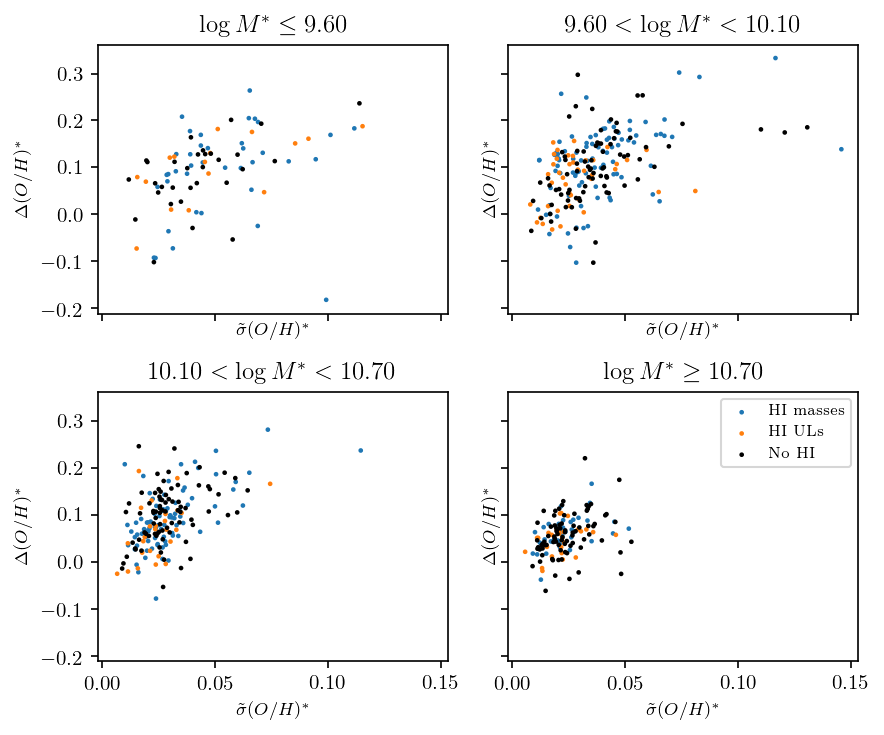}
    \caption{As Figure \ref{fig:hifrac_dec01_subplotmstar}, but relating dispersion of gas-phase metallicity at fixed radius (abscissa) with the strength of metallicity decrement (ordinate), when galaxies are separated by their total stellar mass. Within individual stellar-mass bins, there is a visual correlation between \metdisp and \metdec (see Table \ref{tab:dec-corr_mstarsep} for Kendall's $\tau$ correlation coefficients accounting for upper-limits). This is true for galaxies with measured HI masses (blue), HI upper-limits (orange), and with no measured HI at all (black). The correlation is most striking in the three lower-mass bins, and is marginal for $M^* > 10^{10.7} {\rm M_{\odot}}$.}
    \label{fig:madoh1_dec01_subplotmstar}
\end{figure*}

\begin{figure*}
    \centering
    \includegraphics[width=\textwidth]{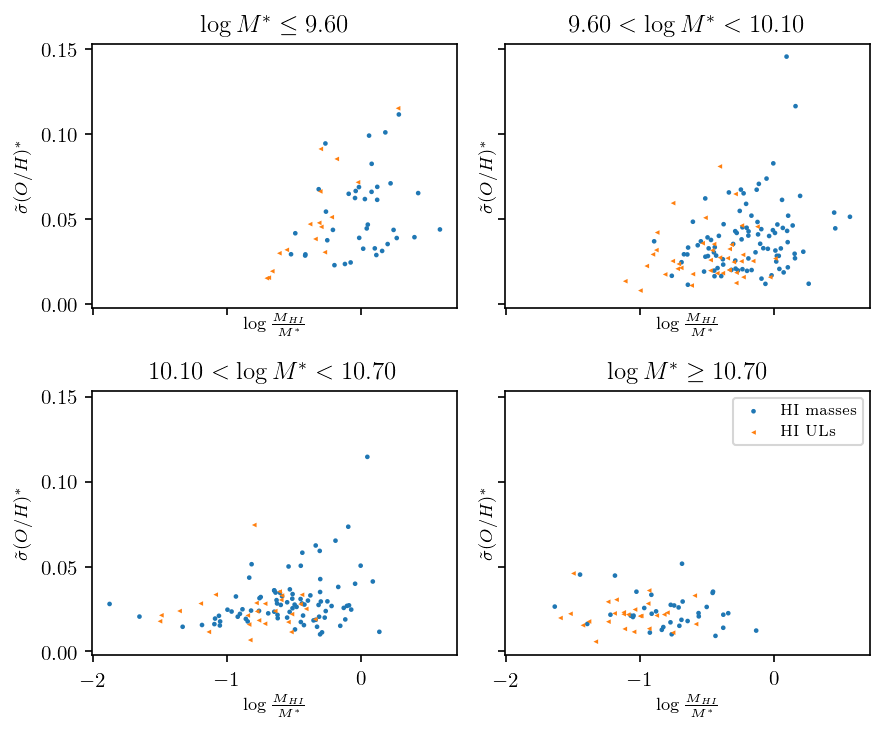}
    \caption{As Figures \ref{fig:hifrac_dec01_subplotmstar} \& \ref{fig:madoh1_dec01_subplotmstar}, but relating dispersion of gas-phase metallicity at fixed radius (abscissa) and $\log \frac{M_{HI}}{M^*}$. Positive HI detections are shown as blue circles, and upper-limits as left-facing, orange triangles. As with \metdec, there is a visual correlation between \& $\log \frac{M_{HI}}{M^*}$ and \metdisp. Like in Figure \ref{fig:hifrac_dec01_subplotmstar} the correlation shown here largely vanishes at $\logmstar > 10.7$.}
    \label{fig:madoh1_hifrac_subplotmstar}
\end{figure*}

Since gas-phase metallicity is a manifestation of a galaxy's (or a region's) chemical evolution, an abnormal metallicity decrement could indicate that an inflow has disturbed the galaxy's abundance profile: such a galaxy might be in an abnormal or non-steady state, with oxygen abundance varying azimuthally as well as radially. We therefore now consider the width of the gas-phase metallicity distribution within the outer radial bin ($1.25-1.75 ~ R_e$), \metdisp, defined as the median absolute deviation distribution of spaxel oxygen abundances within that radial interval. Since \metdisp signifies heterogeneity in chemical enrichment within the outermost radial bin, it is a fair proxy for azimuthal variations in gas-phase metallicity at fixed radius. Figure \ref{fig:madoh1_dec01_subplotmstar} shows the relationship between \metdisp and \metdec in the same four stellar-mass bins used in Figure \ref{fig:hifrac_dec01_subplotmstar}; and Figure \ref{fig:madoh1_hifrac_subplotmstar} shows the relationship between $\log M_{HI} / M^*$ and \metdisp. There is indeed a positive correlation between a strong radial decrement and large metallicity dispersion within a single radial bin; and also between gas-richness \& \metdisp. This implies a close relationship between gas content and non-uniform chemical evolution across a single galaxy.

Table \ref{tab:dec-corr_mstarsep} catalogs the Kendall's $\tau$ rank correlation coefficients between \metdec, \metdisp, and \hifrac. Since the HI fractions include some upper-limits, they cannot be strictly rank-ordered, and so a modified version of Kendall's $\tau$ is used, which accounts for censored data \citep{akritas_murphy_lavalley_95, akritas_siebert_96, helsel_05_nondetects}. As with Figures \ref{fig:hifrac_dec01_subplotmstar}, \ref{fig:madoh1_dec01_subplotmstar}, and \ref{fig:madoh1_hifrac_subplotmstar}, the correlations are measured in four bins of total stellar mass. The final three rows of Table \ref{tab:dec-corr_mstarsep} show the correlation between \metdec and \metdisp for subsamples of the data with HI mass measurements, upper limits, and no data at all.

\begin{table*}[]
    \centering
    \begin{tabular}{c|c|c|c|c|c}
        (1) & (2) & (3) & (4) & (5) & (6) \\
        Correlation & \hi & $\logmstar \le 9.6$ & $9.6 \le \logmstar < 10.1$ & $10.1 \le \logmstar < 10.7$ & $\logmstar \ge 10.7$ \\
            &     & CC (p) & CC (p) & CC (p) & CC (p) \\ \hline \hline
        \hifrac--\metdec & obs. & 0.160 (8.70e-2) & 0.200 (6.15e-4) & 0.275 (1.61e-5) & 0.131 (1.25e-1) \\ \hline
        \hifrac--\metdisp & obs. & 0.177 (5.85e-2) & 0.209 (3.37e-4) & 0.162 (1.08e-2) & -0.128 (8.85e-1) \\ \hline
        \metdisp--\metdec & obs. & 0.368 (6.24e-7) & 0.350 (2.60e-14) & 0.346 (1.16e-12) & 0.281 (5.72e-7) \\ \hline
        \metdisp--\metdec & det. & 0.311 (6.13e-3) & 0.354 (6.08e-7) & 0.402 (4.18e-8) & 0.406 (5.14e-4) \\ \hline
        \metdisp--\metdec & UL & 0.533 (4.56e-3) & 0.234 (3.19e-2) & 0.335 (1.72e-2) & 0.236 (8.74e-2) \\ \hline
        \metdisp--\metdec & None & 0.354 (5.31e-3) & 0.392 (2.18e-7) & 0.268 (4.71e-4) & 0.253 (8.13e-4) \\ \hline
    \end{tabular}
    \caption{Correlation coefficients (p-values) between \metdec, \metdisp, and \hifrac, separated by total galaxy stellar-mass. \textbf{Column (1)} indicates the two quantities correlated, \textbf{column (2)} whether only the subsample with HI observations (obs.), detections (det.), or upper-limits (UL) were used (``None" indicates all galaxies were included---even non-HI-targets), and \textbf{columns (3)-(6)} which bin of total stellar mass is being aggregated. In (3)-(6), the Kendall's $\tau$ rank correlation coefficient is listed, with the p-value in parentheses.}
    \label{tab:dec-corr_mstarsep}
\end{table*}

The correlations between \metdisp and \metdec exist across all bins of total stellar mass, but those involving HI are both strongest and most statistically robust in the intermediate two bins ($9.6 \le \logmstar < 10.7$). At the extremes of stellar-mass ($\logmstar \le 9.6$ and $\logmstar \ge 10.7$), the data cannot establish a correlation with a significance better than 5\%. The lowest-mass bin has very few galaxies, and unlike at high mass, the correlations between \metdec and \metdisp appear consistent with the intermediate-mass bins. In the intermediate mass bins (where the correlations between HI and chemical signatures are significant), the correlations between \metdec and \metdisp are strongest for galaxies with HI detected: this may signify that the effect at play acts on the most HI-rich galaxies at fixed stellar mass. Therefore, we must conclude that the correlation between \metdec, \metdisp, and \hifrac is truly tripartite. 

The largest components of the MaNGA Main Sample, called ``Primary" and ``Secondary", have different distributions in total stellar-mass, redshift, and angular size. This means that at fixed mass the Secondary Sample (higher redshift) will experience PSF-induced blurring at a larger physical scale than will the Primary Sample (lower redshift). Depending on the fundamental scale of the metallicity variations in galaxies, an observer's ability to detect gradients and more local metallicity variations may vary with redshift, possibly giving rise to different distributions of \metdec and \metdisp between Samples\footnote{This effect ought to be strongest at low mass, where the Primary Sample targets are a factor of several closer than Secondary Sample targets \citep{manga_sample_wake_17}}. To test this effect, we calculate the correlation between \metdec and \metdisp in bins of mass \emph{and} separated by Primary or Secondary Sample. For $\logmstar < 9.6$, the Primary Sample (56 galaxies) strongly dominates in number over the Secondary (8 galaxies), and in the Secondary Sample we are unable to establish a correlation at 5\% significance; at $\logmstar \ge 10.7$, a significant correlation only manifests in the Primary Sample. Though we do not show them, the \metdec-\metdisp correlations are very similar in magnitude between Primary \& Secondary Samples in the intermediate mass bins. We are satisfied, though, that the observed effects are not completely confined to either Primary or Secondary Sample---that is, that redshift effects are not dominant.

The trends noted here do not manifest as a result of contamination of the GBT beam by a secondary, nearby source: after eliminating galaxies with a companion within one beam HWHM (as cataloged in the MaNGA-GEMA VAC---\citealt{argudo-fernandez_2015_environment}), the observed trends persist. The observed abundance dispersions in this MaNGA sample have a mode qualitatively in line with those found in a sample of PHANGS galaxies \citep{kreckel_2019_phangs_metgrads}: the PHANGS study relied on the sulfur-based \texttt{PG16-S2} calibration, but the two have been shown to be nearly identical \citep{pilyugin_grebel_2016}.

\subsection{Accounting for metallicity measurement uncertainties}
\label{subsec:met_unc}
The correlations noted above have neglected the varying uncertainties of the measured spaxel metallicities. Though the data-quality cuts imposed in Section \ref{subsubsec:data_quality} eliminate the spaxels with the very highest uncertainty, the non-uniform signal-to-noise ratio of the measured lines should be accounted for when estimating metallicity's spatial variation (particularly azimuthal). One additional measurements of azimuthal metallicity variation is employed: normalized excess variance (NXS) is used commonly in the analysis of quasar variability \citep{nandra_97_agn-variability, vaughan_03_xray-variability}, and intuitively compares the dispersion in an empirical distribution with each measurement's uncertainty. To replace \metdisp, we adopt a modified version of NXS, which accounts for the spatial covariance in the spaxel metallicity estimates.

\subsubsection{An emergent uncertainty-decrement correlation}
\label{subsubsec:indicator_sn}
The effects of uncertainty in emission line flux measurements are not felt equally across all metallicities and radii. The \texttt{PG16-R2} metallicity indicator is constructed such that [NII] lines determine the branch of $R_{23}$ selected. This works well at the higher metallicities typically encountered in high-mass galaxies and in galaxy centers; at lower metallicities, nitrogen lines become fainter. In addition, the curvature of the \texttt{PG16-R2} indicator in $(O/H)^*$--$\log R_{23}$ space at oxygen abundances less than about 8.4 makes the strong-line metallicity very sensitive to the value of $R_{23}$. These effects combine to starkly increase the dispersion of strong-line metallicity estimates in galaxy outskirts, relative to the relatively more metal-rich interiors, at the typical S/N ($\sim 10$) conditions of IFU surveys; and could somewhat enhance observed metallicity decrements relative to other strong-line calibrations. This work's use of median-statistics in computing \metdec and \metdisp alleviates the impact of errant, low-metallicity outliers in this regime, but the effect may be important in explaining some portion of the correlation between those two quantities.

\subsubsection{Spatial covariance and its effects}

The resolved metallicity estimates described above also include an important, unaccounted-for element of spatial correlation. Failing to account for this, and simply treating each spaxel as independent, severely mis-estimates the intrinsic dispersion in the measurements. \citet{manga_dap} finds that the spatial correlation between nearby spaxels' emission line flux measurements is well-described by a Gaussian with a length-scale of 0.96 arcseconds (1.92 spaxels) for $\lambda = 4700 {\rm \AA}$. Under those spatial correlation conditions, simulated measurements within a single, MaNGA-like radial annulus have their observed dispersion significantly depressed (by about an order of magnitude). A separate, geostatistical kriging analysis of the spaxels in a randomly-selected subset of galaxies reveals that under a Gaussian (RBF) spatial covariance model, the best-fit length-scale for the metallicity covariance is somewhat smaller, approximately 0.8 arcseconds (1.59 spaxels).

Spatial covariance in the measured spaxel metallicities would be less problematic if all metallicity maps obtained from the strong-line calibrations were equally dense in the plane of the sky; however, in adopting relatively stringent spaxel-wise data-quality cuts, we have imposed unequal spatial sampling from galaxy to galaxy, which likewise impacts the sampling of the spatial PSF. The typical definition of NXS uses the ``average variance" of the data points to characterize the uncertainty of measured values: in the case of covariate data, this quantity is the one that overestimates the true spaxel-to-spaxel variance. This quantity is therefore replaced by $|\Sigma|^{1/k}$ (the determinant of the spatial covariance matrix, raised to the power of the reciprocal of the number of measurements). $|\Sigma|^{1/k}$ qualitatively describes the combined effects of intrinsic and covariate uncertainty \citep{dumbgen_tyler_05, paindeveine_08}. We denote the excess variance beyond the spatially-covariate metallicity measurement uncertainties as \cnxs.

\subsubsection{Stellar-continuum-induced spatial covariance}
Though the explicit spatial PSF of the MaNGA observations appears to be the dominant form of spatial covariance, it is almost certainly not the sole form. Using the best-fit covariance lengthscale from the kriging model, a sizable fraction of MaNGA galaxies in our sample exhibit \emph{negative} values of \cnxs. In other words, the spatial covariance model appears to \emph{overestimate} the observed dispersion in the spaxel metallicity measurements (i.e., the metallicity measurements are under-dispersed relative to the model). This behavior is not alleviated by reasonable contractions of spaxel-wise metallicity measurements' uncertainties (which modulate the spatial correlation term, and could, if overestimated, bias variance estimates high). This is true as a general rule across all bins of galaxy stellar mass, on average; and though there is some diversity in \cnxs value within the sample, the driver for this diversity is difficult to isolate. To wit, negative values of \cnxs do not appear to correspond closely with intrinsic galaxy properties, such as stellar mass or placement in the Primary or Secondary MaNGA sample; nor observational conditions, such as redshift (inadequate telluric absorption correction might produce abberant emission-line uncertainties, which would propagate explicitly into the covariance model); nor IFU size.

The observational under-dispersion in metallicity measurements, at its core, reflects an under-dispersion in the emission line flux measurements made by the MaNGA DAP. Therefore, we may also look to the DAP for possible origins of this effect. A plausible candidate explanation is the DAP's ``hybrid" spatial binning scheme: in an effort to produce high-quality measurements of stellar kinematics, the DAP aggregates nearby spaxels into Voronoi cells, until a target $(S/N)_{g}$ of 10 is reached \citep[Section 6.3]{manga_dap}. The best-fit to the stellar populations of the aggregate spectrum is then used to initialize joint stellar-gas fits for individual constituent spaxels, with the spaxels' stellar kinematics \emph{fixed} to the values from the aggregate fit. Though this hybrid approach does ensure robust fits to the stellar kinematics, it has the potential to induce modelling deficiencies which propagate into measured emission line fluxes. Thus, adopting the kinematics of a global best-fit might induce a systematic when measuring gaseous emission in individual spaxels. ``Global-versus-local" failures of this type have well-established effects on photometric data \citep{zibetti_2009, sorba_sawicki_15}; and similar effects have also been shown to bias measurements of, for example, stellar mass-to-light ratio made using optical spectra \citep[Section 4.3]{pace_19b_pca}. Based on analysis of a related failure mode (stellar model library mismatch: \citealt{belfiore_2019_mangadap-emissionline}), it is plausible that emission lines located within or near stellar absorption features (Balmer lines and [NII]) could be mis-estimated by as much as 0.1 dex. Depending on the diversity of stellar kinematics and surface brightnesses within on Voronoi cell, such mis-estimates could be spatially correlated; and thus, derived metallicity estimated would also be correlated.

Unfortunately, it is not straightforward to measure or correct these spatial correlations, since the size of the Voronoi cells varies within a galaxy. There is no convincing correspondence between galaxies with strongly-negative \cnxs and those with low median $(S/N)_{g}$; nor with low integrated Balmer or [NII] flux; nor with low integrated star-formation rate. Galaxy-integrated measurements, though are not the best measurement of local variation; and so we recommend a more exacting, aggregated analysis of emission line flux measurements with respect to stellar continuum residuals. With these concerns in mind, but without a well-motivated correction to the spatial correlation model, we adopt \cnxs as the measurement of metallicity dispersion within a single galaxy.

\subsubsection{Population correlations with metallicity measurement uncertainties}

We now move to correlating \cnxs (as was done above using \metdisp) against \hifrac and \metdec. Table \ref{tab:dec-corr_mstarsep-cnxs} reports censored rank correlation coefficients computed in the same manner as in Table \ref{tab:dec-corr_mstarsep}. For the sake of brevity, plots replicating Figures \ref{fig:hifrac_dec01_subplotmstar}, \ref{fig:madoh1_dec01_subplotmstar}, and \ref{fig:madoh1_hifrac_subplotmstar} have been omitted; but they show similar behavior to the trends involving \metdisp. Indeed, the principal results of a positive correlation between metallicity dispersion within one radial bin, large \hifrac, and steep metallicity decrement, are largely preserved when accounting for spaxel metallicity measurement uncertainty and the associated spatial covariance. The only salient difference arising from the integration of spaxel uncertainty is a lessening of the correlation significance for the higher-stellar-mass bins ($10.1 \le \logmstar < 10.7$ and $\logmstar \ge 10.7$). As a general rule, there are no significant correlations between \cnxs, \hifrac, and \metdec at high galaxy total stellar mass, indicating that the spaxel metallicity variance and covariance corrections are more important in that regime. The persistence of the correlations at $\logmstar < 10.1$ indicates that at a population level, gas content and metallicity variations track together. Thus, they may arise from just one evolutionary mechanism.

\begin{table*}[]
    \centering
    \begin{tabular}{c|c|c|c|c|c}
        (1) & (2) & (3) & (4) & (5) & (6) \\
        Correlation & \hi & $\logmstar \le 9.6$ & $9.6 \le \logmstar < 10.1$ & $10.1 \le \logmstar < 10.7$ & $\logmstar \ge 10.7$ \\
            &     & CC (p) & CC (p) & CC (p) & CC (p) \\ \hline \hline
        \hifrac--\cnxs & obs. & 0.202 (3.06e-2) & 0.131 (2.47e-2) & 0.0173 (7.97e-1) & -0.0681 (4.29e-1) \\ \hline
        \cnxs--\metdec & obs. & 0.292 (7.30e-5) & 0.284 (6.31e-10) & 0.0859 (7.78e-2) & -0.0138 (8.07e-1) \\ \hline
        \cnxs--\metdec & det. & 0.198 (8.27e-2) & 0.233 (1.00e-3) & 0.140 (5.62e-2) & 0.0825 (4.87e-1) \\ \hline
        \cnxs--\metdec & UL & 0.633 (7.34e-4) & 0.161 (1.41e-1) & 0.102 (4.81e-1) & 0.0143 (9.33e-1) \\ \hline
        \cnxs--\metdec & None & 0.260 (4.14e-2) & 0.368 (1.19e-6) & 0.0250 (7.48e-1) & -0.0778 (3.06e-1) \\ \hline
    \end{tabular}
    \caption{As Table \ref{tab:dec-corr_mstarsep}, but replacing \metdisp with \cnxs, a measure of excess variance relative to a spatial covariance model.}
    \label{tab:dec-corr_mstarsep-cnxs}
\end{table*}

\section{Inflow model: does it explain the metallicities?}
\label{sec:modeling}

We consider now whether an inflow falling onto galaxy outskirts could give rise to the chemical inhomogeneities we observe there. In this scenario, the low-metallicity inflow would mix with and ``dilute" the ambient ISM, decreasing the measured metallicity. Since strong decrements and high HI fractions are also associated with an increased metallicity dispersion at fixed radius, it is clear that the inflow is not incident on the entire galaxy, or at least is not equally well-mixed with the ambient ISM. That is, the inflow covering fraction $f_c$ as it manifests chemically is somewhat less than unity. Adopting a simplistic model of dilution, where the ambient metallicity at some radius is diluted locally by an additive factor $d_{in}$, the observed local metallicity will be given by
\begin{equation}
    \metlin_{obs} = \frac{\metlin_{amb} + \metlin_{in} ~ d_{in}}{1 + d_{in}}
\end{equation}
where $\metlin_{amb}$ and $\metlin_{in}$ are the oxygen abundance by number, in linear units, in the ambient ISM (before the inflow) and the inflowing gas. In other words, $\metlin = 10^{\met - 12}$.
By arithmetic rearrangement, we obtain
\begin{equation}
    d_{in} = \frac{\metlin_{amb} - \metlin_{obs}}{\metlin_{amb} - \metlin_{in}}
\end{equation}
a positive, finite quantity under the restriction $\metlin_{amb} > \metlin_{in}$.

\begin{figure*}
    \centering
    \includegraphics{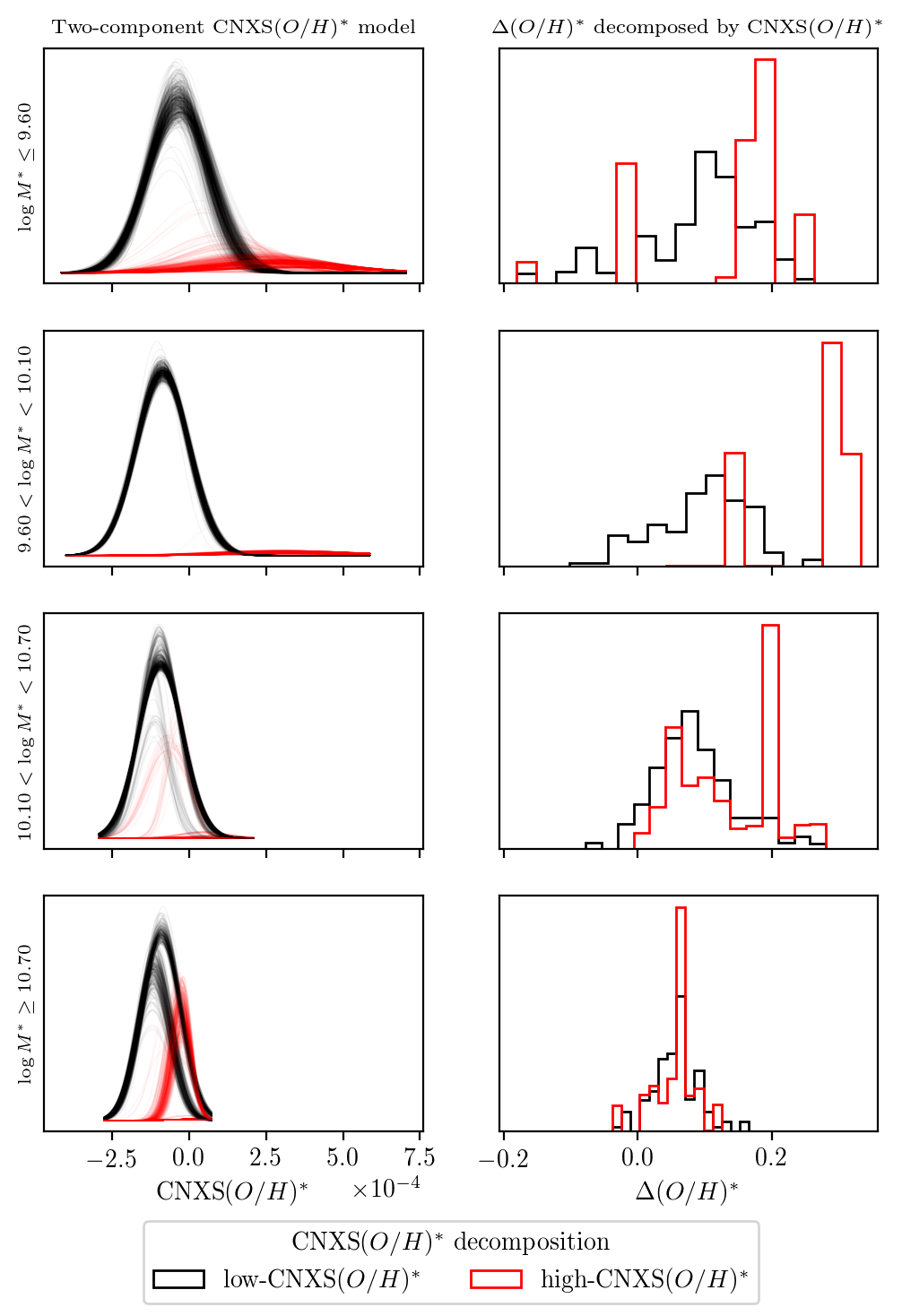}
    \caption{Decompositions of the $M^*$-separated sample of galaxies on the basis of their covariance-adjusted metallicity dispersion (\cnxs) in the interval $1.25-1.75 ~ R_e$. Each row shows the decomposition results for one bin of total galaxy stellar mass. In the left-hand column are shown five-hundred two-component decompositions of \cnxs, under a bootstrap-resampling approach (80\% of points retained). The black (red) curves show the best-fit distributions of the low- (high-) -\cnxs populations. In the right-hand panel are shown the empirical distributions of \metdec for the low- and high-\cnxs populations (black and red), weighted by how often a galaxy is sorted into each population. Both populations appear to have relatively consistent modes across bins of stellar mass, though the low-\cnxs population is visually much more prominent at low stellar mass. Strikingly, the members of the high-\cnxs population tend to have high values of \metdec in the low- and intermediate-mass bins.}
    \label{fig:decomposition}
\end{figure*}

Using this rough, but intuitive approximation, we next aim to constrain the permitted values of the inflow dilution factor $d_{in}$ as a function of total stellar mass. To determine the ambient metallicity, we seek to separate out galaxies with substantially-higher-dispersion metallicity measurements than typical for a particular mass range. Within a given stellar-mass bin, we decompose the observed distribution of $\cnxs$ into two components, according to a Dirichlet-process Gaussian mixture model (DPGMM) as implemented in \texttt{scikit-learn} \citep{scikit-learn} (see Figure \ref{fig:decomposition}). The component with the smaller metallicity dispersion is taken as the fiducial, undiluted sample, used to characterize the ambient metallicity; and the component with the larger mean metallicity dispersion is taken as a comparison sample with unknown dilution characteristics. We elect to decompose on $\cnxs$ because it allows easier population-statistics to be made with HI mass fraction and metallicity decrements within the fiducial and comparison samples.

Each stellar mass bin is boostrap-resampled (80\% of points retained) a total of 500 times, resulting in a set of unique reclassifications into low-\cnxs and high-\cnxs populations. For the two lowest-mass bins ($\logmstar < 10.1$), the decomposition is stable, and the presence of two components is strongly favored; whereas for stellar masses greater than $10^{10.1} {\rm M_{\odot}}$, the presence of a high-\cnxs population is only marginally indicated by the data. From this, one might infer that diluting inflows are relatively rare in incidence at such masses. In the lower-mass bins ($\logmstar < 10.1$), the decomposition indicates that 10\%-25\% of galaxies host a diluting inflow at any given time. This is similar to the $\sim 25\%$ estimate of \citet{hwang_2019_manga_almrs}. A decomposition on \metdec (not shown) yields equivalent results, with the exception of a more reliable two-component decomposition in the second-highest bin of stellar mass ($10.1 \le \logmstar < 10.7$). Within each bin of stellar mass, there is no discernable bias in total stellar mass between the fiducial (low-\metdisp) and comparison (high-\metdisp) populations. There are, however, differences in \metdec and \hifrac between populations. For example, we note that the high-\cnxs populations have median HI mass fraction ($\frac{M_{HI}}{M^*}$) between 0.15 dex and 0.25 dex larger than the low-\cnxs populations (considering only galaxies with positive HI detections). In the low- and intermediate-mass bins, there is also a difference ($0.05-0.2 ~ {\rm dex}$) between the low- and high-\cnxs populations' distributions of \metdec (Figure \ref{fig:decomposition}, right-hand panel). These results are reminiscent of the trends found in Section \ref{sec:trends}.

Within each stellar mass bin, the comparison (high-\metdisp) sample's median oxygen abundance in radial bin 1 ($\met_{comp,1}$) is taken as the observed, post-inflow metallicity ($\met_{obs}$). A range of possible pre-inflow metallicities ($\met_{in}$) are used, with the following limits
\begin{itemize}
    \item The median oxygen abundance in radial bin 1 of the fiducial sample ($\met_{fid,1}$)
    \item The average of $\met_{comp,1}$ and $\met_{fid,1}$
\end{itemize}

The metallicity of the accreted gas is set to one of two values obtained through a gas-regulator-model analysis of the same type as \citet{schaefer_19_accretion}, but using the \texttt{PG16-R2} metallicity calibrator: a value of 7.24 ($\sim 1/30 Z_{\odot}$) corresponds to the mean accreted metallicity of satellites of low-mass ($\logmstar < 10$) galaxies; and a value of 7.52 ($\sim 1/15 Z_{\odot}$) corresponds to the mean accreted metallicity of satellites of high-mass ($\logmstar > 10.5$) galaxies. While these metallicities by definition pertain to accretion by low-mass galaxies ($9 < \logmstar < 10$), and are therefore most appropriate to the two lowest-mass bins, they are realistic order-of-magnitude estimates. An alternative would be to adopt for higher-mass galaxies the typical metallicities of Milky Way high-velocity clouds (HVCs), which are somewhat sub-solar \citep{wakker_2004_hvc_bookchapter}, though not universally external in origin \citep{fox_2016_smithcloud}. We do not do show results for typical HVC metallicities, since the model is relatively insensitive to inflow metallicity.

\begin{figure*}
    \centering
    \includegraphics[width=\textwidth]{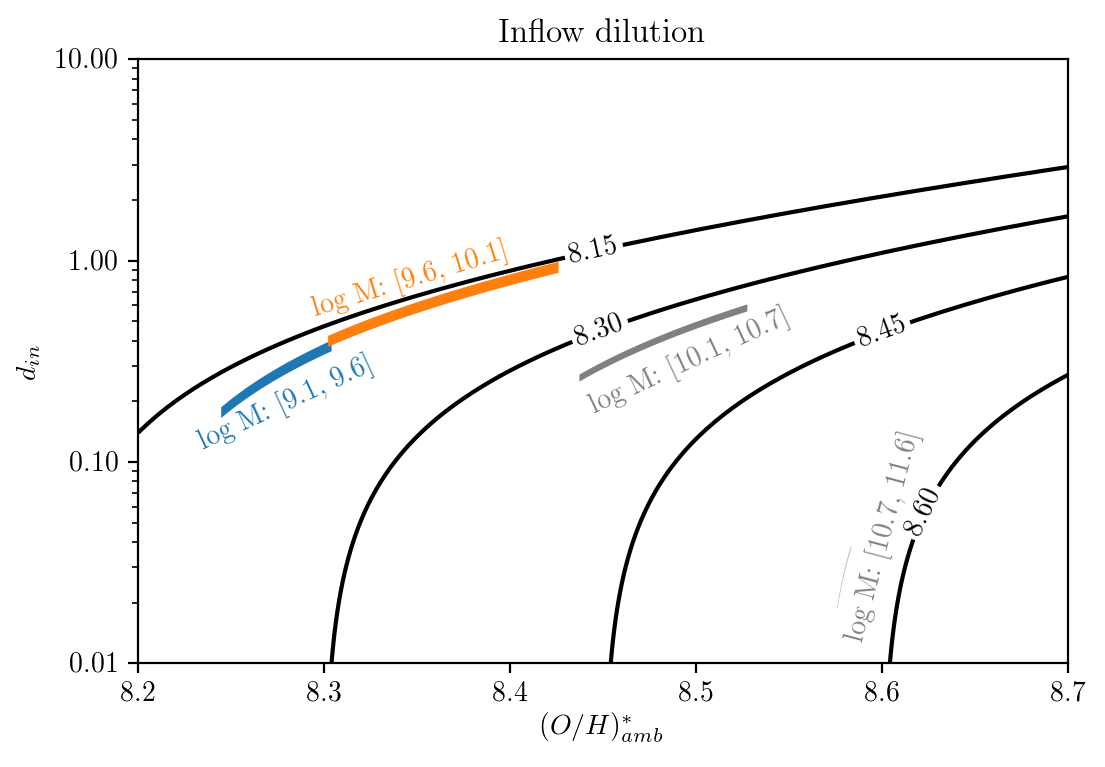}
    \caption{A toy model of inflow's effect locally on observed metallicity, assuming a range of ambient \& observed (pre-inflow \& post-inflow) metallicities, along with a range of inflow metallicities taken from a best-fit equilibrium model. A covariate range of local dilution factor and ambient metallicity (dependent on the choice of inflow metallicity) is shown for each of the four mass bins: the two lowest-mass bins are shown in color, since the two-component GMM used to estimate their pre- \& post-inflow metallicities is fairly well-characterized; the two highest-mass bins are shown in gray, since the two-component decomposition is less reliable in that regime. Overlaid are contours representing the corresponding, average observed gas-phase metallicity for a given combination of $\met_{amb}$ and $d_{in}$.}
    \label{fig:oh-ambient_d-in_color-oh-obs}
\end{figure*}

$d_{in}$ is computed for each combination of $\met_{amb}$ and $\met_{in}$. The results are shown in Figure \ref{fig:oh-ambient_d-in_color-oh-obs}. We find that according to this picture of dilution by lower-metallicity gaseous inflows, local star-forming gas reservoirs encounter inflows with between 10\% and 90\% of ambient gas surface-density. The most massive, star-forming galaxies ($> 10^{10.7} {\rm M_{\odot}}$) seem to be somewhat distinct in that any inflow is constrained to be relatively small in magnitude compared with their existing local gas reservoirs; at high stellar mass, the multi-component decomposition of \cnxs does not heavily favor a high-\cnxs population, so this result is much less certain. In contrast, our simple model permits lower-mass galaxies to experience significant inflows ($20-90\%$) relative to their existing local gas reservoirs.

The assumed ambient metallicity can be thought of as a reflection of how commonly inflows are incident at local scales upon a star-forming disk: if most regions are subject to more constant refueling (i.e. dilution), then the ambient metallicity is implied to be higher. That is, when only the most metal-rich tail of the metallicity distribution reflects the galaxy's recent star formation free from inflows, the local dilution must be stronger on average. This implies a large inflow covering fraction, so we evaluate the higher inflow dilution factors as somewhat less likely, given the large observed metallicity dispersion at fixed radius.

\subsection{Total \hi, dust, and extended UV disks in the accretion-dilution model}
\label{subsec:hifrac_diversity}

Under the toy model described above, in which the ambient ISM of galaxies is diluted by lower-metallicity gas with extragalactic origin, there should be a correspondence between the degrees of dilution and local H\textsc{i}-enhancement. In Section \ref{sec:modeling}, we computed that at total stellar masses of approximately $10^{9.6}-10^{10.1} M_{\odot}$ and accreted-gas metallicities of 7.24, local dilution factors are implied to be approximately 50\%, with a covering fraction somewhat less than one. Within individual stellar-mass bins, the higher-$\tilde{\sigma}(O/H)^*_1$ population has a larger HI mass fraction on average, by between 0.15 dex and 0.25 dex in the stellar mass intervals $ \logmstar < 9.6$ and $9.6 \le \logmstar < 10.1$.

To evaluate the possible inflow-induced enhancements in H\textsc{i} mass associated with the metallicity dilutions, we adopt a plausible, if simplistic description of a galaxy's H\textsc{i} radial mass profile: an exponential disk \citep{bigiel_blitz_12-gasprofile} with an inner core (representing a transition to predominantly molecular gas) and an outer cutoff. Interior to an atomic-to-molecular transition radius $r_t$, the atomic gas mass surface density is constant, accounting for a possible saturation of atomic gas content and a transition to a molecular-dominated core \citep{krumholz_2009_hi-h2, schruba_2018-hi}. Between $r_t$ and a cutoff radius $r_c$, the radial profile is an exponential, with a scale-length of $r_t$; and exterior to $r_c$, there is no gas. 

As an illustrative example, we consider the case of a hypothetical low-mass galaxy, with $M^* \sim 10^{9.85} {\rm M_{\odot}}$ and $\log \frac{M_{HI}}{M^*} \sim -0.2$ (the median of the low-\cnxs population for that stellar-mass bin). Following roughly a set of HI mass-size relations \citep[as collated and expanded in \citealt{blue-bird_davis_2020}]{verheijen_sancisi_2001_hi-mass-size, wang_koribalski_2016_hi-mass-size, diskmass_x_martinsson_2016}, we estimate a physical HI disk cutoff radius of $\sim 20 {\rm kpc}$ (or roughly $8 R_e$). We rely on previous calibrations of bandpass-specific scale-lengths with respect to $r_{25}$ to estimate the dimensionless ratio $\frac{<h_{\Sigma_{tot~gas}} / r_{25}>}{<h_{0.7 \mu m} / r_{25}>} \approx 1.6$ \citep[see][Table 7]{casasola_2017_dustpedia-scalelengths}. In units of $R_e$, this ratio is adopted as $r_t$.

We will briefly use this model profile as a fiducial for a galaxy without an inflow, and evaluate possible inflow-related enhancements to it. Guided by our dilution model, we assume that a gaseous inflow provides a 50\% enhancement to local star-forming gas reservoirs, with a covering fraction of 50\%. Under this fiducial model, the mass fraction contained in the radial interval [$1.25 ~ R_e$, $1.75 ~ R_e$] is approximately 6\%. Consequently, an inflow with the characteristics described above would affect a minuscule ($\sim 1\%$, or $< 0.01 ~ {\rm dex}$) enhancement in the \emph{global} gas mass. This admittedly-limiting case is substantially less than typical differences between HI fractions in low- and high-\cnxs populations (0.15 - 0.25 dex). In contrast, an inflow affecting all radii between $1.25 R_e$ and $r_c$ would enhance a galaxy's global HI mass by $\sim 0.1 ~ {\rm dex}$ (assuming the same covering fraction and local gas enhancement). This is more plausible, but still does not match the HI differential that the \cnxs decomposition suggests. As the assumed outer extent of the inflow grows to exceed $r_c$, the enhancement in total gas mass becomes insensitive to $r_c$; in order to reach the large HI fractions of the high-\cnxs population, the inflow's radial profile must become shallower (i.e., the scale-length must increase). For example, an inflow profile incident only on $r > 1.25 ~ R_e$, with a 50\% local mass enhancement at $1.25 ~ R_e$, $r_t = 2 ~ R_e$ (a shallower slope than the fiducial), and $r_c = 16 ~ R_e$ (a larger cutoff radius than the fiducial) achieves a global gas mass enhancement of nearly $0.2 ~ {\rm dex}$. Modulating the inflow covering fraction with respect to radius is qualitatively similar to adopting a longer inflow scale-length: if this were the case, metallicity profiles would steepen when measured at larger radius, but \cnxs would lessen (as the metallicity distribution function at fixed radius becomes unimodal). These examples illustrate that if an inflow is responsible for the abundance trends found in this work, there ought to be a \emph{significant} gaseous component at large galactocentric radius.

An extended gaseous tail could also form stars, albeit with lower efficiency \citep{rafelski_2016}. Indeed, a star-forming ``plume" has been reported around M101 \citep{mihos_2013_m101-uvdisk}, as well as associated with MaNGA ALM-region candidates \citep{hwang_2019_manga_almrs}. That said, no statistical enhancement of NUV-to-$r$-band Petrosian radius ratio (which might signal low-efficiency star formation in the inflowing gas) is found in high-\cnxs galaxies compared to peer galaxies at similar mass.

Inflows, if present, may also have a dust component, since their metallicity is strongly constrained to be nonzero. So, for an inflow with large enough mass and dust-to-gas ratio, the stellar populations may appear to have a greater attenuating dust foreground with respect to the typical galaxy without an inflow. To investigate this possibility, we adopt the importance-sampling inference method of \citet{pace_19a_pca}, and use a representative library of 40,000 synthetic ``training" SFHs and their spectra to infer the foreground V-band optical depth arising from a two-component dust model \citep{charlot_fall_00}. However, we find that in the radial interval $1.25-1.75 ~ R_e$, the inferred V-band optical depth (arising from both the diffuse and dense dust components) is not noticeably enhanced in galaxies inferred to belong to the high-\cnxs population, at fixed stellar mass. This lack of visible foreground dust enhancement could be explained by a few factors: first, the number of high-\cnxs galaxies in each mass bin is small by construction. Second, the geometry of the inflow with respect to the line of sight is unknown, and an inflow may not produce an attenuation signature along all lines of sight. Finally, at the expected inflow metallicities, the dust-to-gas ratio is likely a factor of several smaller than at ambient metallicities \citep{kahre_walterbos_2018_legus_dusttogas}. Thus, the local dilution observed may not produce a detectable dust attenuation signature.

%% file: discussion.tex
\section{Discussion and Conclusions}
\label{sec:disc}

Galaxies are widely understood to exchange gas with their surroundings. In this work, we relate the abundance patterns present within individual MaNGA star-forming galaxies to their total atomic gas content. We find
\begin{itemize}
    \item Galaxies with high H\textsc{i} masses relative to their total stellar-masses tend to possess strong gas-phase metallicity decrements between their innermost star-forming regions ($0.0-0.5~R_e$) and those at slightly larger radii ($1.25-1.75~R_e$). This is similar to an effect noted in a sample of DustPedia galaxies \citep{devis_2019_dustpedia}.
    \item Those same galaxies also tend to have relatively wide distributions ($> 0.05~{\rm dex}$) of gas-phase metallicity in the radial interval $1.25-1.75 ~ R_e$, compared with both peer galaxies of similar mass and a separate sample of star-forming galaxies at higher spatial resolution, but with metallicities obtained using the same strong-line calibration \citep{kreckel_2019_phangs_metgrads}.
    \item These effects are limited to relatively low-mass galaxies: galaxies with $\logmstar > 10.7$ seem to lack both strong metallicity decrements and large metallicity dispersions within the interval $1.25-1.75 ~ R_e$; and galaxies at slightly lower stellar mass ($10.1 \le \logmstar < 10.7$) have weak correlations at the population level between metallicity dispersion, radial decrement strength, and HI content. This indicates that inflows occur less frequently in this regime, or have a smaller overall impact. The small number of galaxies with measured HI in the lowest-mass bin coincides with somewhat less significant correlations involving \hifrac, but the magnitude of the correlations between \metdec, \metdisp, and \cnxs do not diminish commensurately. More HI observations of low-mass galaxies may result in more robust correlations with HI.
    \item Galaxies with abnormally-large metallicity dispersion between 1.5 and 2.5 $R_e$ have on average $0.15-0.25 ~ {\rm dex}$ more HI when normalized by total stellar mass.
\end{itemize}

We attempt to explain the combination of effects by invoking azimuthally-asymetric, low- (but not zero-) -metallicity inflows with a strong atomic component. Indeed, \citet{pezzulli_fraternali_2015} used an analytic chemical-evolution framework to predict that inward flows of gas produce steeper-than-normal metallicity gradients. The gas-rich inflows explored in this work are observationally distinct from those brought about by galaxy mergers, which have been shown to \emph{flatten} abundance gradients, rather than steepen them \citep{rupke_2010_inflows_metgrad}. Since enhanced radial metallicity decrements are also associated in our sample with increases in the \emph{metallicity dispersion at fixed radius}, we conclude that any inflow ought to have a significant covering fraction---at least 10\%, in order to noticeably widen the metallicity distribution in one radial interval; but less than 100\%, a case in which \metdec would rise, but \metdisp would not. Under our basic model, the plausible range of accreted ISM metallicities implies a range of inflow dilution factors (the margin by which a \emph{local} gas reservoir's mass is enhanced) of 20\%-90\%. Though a higher-metallicity inflow universally implies a larger inflow for the same pre-inflow ambient and post-inflow observed metallicity, this effect becomes all but negligible at higher stellar masses (higher ambient metallicities).

As discussed by \citet{schaefer_19_accretion}, the presence of a nearby massive halo is associated with accreted gas that is more enriched. This means that even within one mass bin, the metallicity of accreted gas could vary by a factor of several. In other words, the same enhanced metallicity decrement would require a smaller inflow by mass for a satellite of a low-mass galaxy than for a satellite of a high-mass galaxy. Once MaNGA observations and HI follow-up are complete, it may be possible to modulate the assumed inflow metallicity based on an inflow host's environment. Additionally, galaxy mass itself may impact the inflows it experiences: \citet{muratov_2017_FIRE_cgm_metals} finds that a high-mass halo ($M^* \sim 10^{10.8} {\rm M_{\odot}}$) rarely sustains inflows that reach its interior ($< 0.25~R_{vir}$); whereas at lower halo mass ($M^* \sim 10^{9.3} {\rm M_{\odot}}$), inflows have a duty cycle of about 50\%! This agrees qualitatively with the observed dearth of evidence for inflows in the highest-mass bin in this work.

Though the dilution factors obtained according to this simple model imply moderate enhancements in local gas supply, there ought to be a connection to the reservoir of gas supplying the entire galaxy. With the assumed inflow metallicities, the HI mass associated with the dilution effect in strictly the radial interval $1.25-1.75 R_e$ is substantially smaller than the average difference in HI fractions between the high-metallicity-dispersion and low-metallicity-dispersion populations. Preserving the connection between local and global gas supply seems to require a gaseous component with relatively large radial scale-length. The GBT observations (having a FWHM of 8.8') are capable of detecting HI residing far away from the star-forming disk, and any diluting inflows incident upon a galaxy's disk could be understood as a small fraction of gaseous disk that extends outward to $10 R_e$ or beyond \citep{bigiel_blitz_12-gasprofile}.

Alternative explanations to the diluting-inflow hypothesis include intra-galaxy modulation of star formation efficiency (SFE): \citet{schaefer_19_ohno} reports radial variations in SFE of nearly an order of magnitude, but the degree of variance within single galaxies and at constant radius is at present unexplored \citep[see also][]{bigiel_leroy_08}. Galaxy regions with lower SFE may have lower metallicity than peer regions at similar radius. In order to explain the diversity of HI mass fraction at fixed stellar mass, the SFE would have to be coherently depressed across the entire galaxy for a significant portion of the age of the universe, which we consider an unlikely scenario.

\begin{table*}
\centering
\begin{tabular}{|c|c|c|c|c|c|c|c|c|}
(1) & (2) & (3) & (4) & (5) & (6) & (7) & (8) & (9) \\
\texttt{plateifu} & \texttt{mangaid} & $\log M_{HI}$ (or U.L.) & HI meas. type & \logmstar & $z$ & $\log \hifrac$ & \metdisp & \metdec \\ \hline\hline
9501-6101 & 1-384726 & 9.899 & 0 & 9.807 & 0.03924 & 0.09194 & 0.146 & 0.139 \\ \hline
8452-12701 & 1-167678 & nan & 2 & 9.948 & 0.04017 & nan & 0.131 & 0.185 \\ \hline
8323-12704 & 1-405813 & nan & 2 & 9.688 & 0.03819 & nan & 0.121 & 0.175 \\ \hline
8156-12703 & 1-38894 & 9.974 & 0 & 9.815 & 0.04192 & 0.159 & 0.117 & 0.333 \\ \hline
7495-9101 & 12-129610 & 9.443 & 1 & 9.168 & 0.03235 & 0.275 & 0.115 & 0.188 \\ \hline
9506-12704 & 1-299793 & 10.469 & 0 & 10.424 & 0.04884 & 0.0455 & 0.115 & 0.237 \\ \hline
9485-9102 & 1-121994 & nan & 2 & 8.980 & 0.01918 & nan & 0.114 & 0.237 \\ \hline
8146-9101 & 1-556506 & 9.819 & 0 & 9.538 & 0.02393 & 0.281 & 0.112 & 0.183 \\ \hline
8259-9101 & 1-257822 & 9.409 & 0 & 9.230 & 0.01980 & 0.180 & 0.101 & 0.169 \\ \hline
8936-6104 & 1-152828 & 9.419 & 0 & 9.361 & 0.01570 & 0.0584 & 0.0993 & -0.182 \\ \hline
\end{tabular}
\caption{A segment of the machine-readable table aggregating total galaxy stellar masses, chemical variations, and HI masses/upper-limits (where available). \textbf{Columns (1) \& (2)} provide a galaxy's MaNGA-ID and \texttt{plate-ifu} designations, \textbf{column (3)} the HI mass or upper-limit (``nan" if not in the HI follow-up campaign), \textbf{column (4)} the HI measurement type (0 signifies a measurement, 1 an upper-limit, and 2 not targeted), \textbf{column (5)} the total galaxy stellar mass \citep{pace_19b_pca}, \textbf{column (6)} the optical redshift, \textbf{column (7)} the ratio of the HI mass to the stellar mass (\hifrac), \textbf{column (8)} the metallicity dispersion in the radial interval $1.25-1.75 ~ R_e$ (\metdisp), and \textbf{column (9)} the measured metallicity decrement (\metdec).}
\label{tab:galaxies}
\end{table*}

Like in \citet{hwang_2019_manga_almrs}, this study finds evidence for gaseous inflows affecting a sizeable fraction of star-forming galaxies in the nearby universe ($\sim 10\%$), as reflected in the presence of anomalously-low-metallicity, star-forming gas. While \citet{hwang_2019_manga_almrs} defines ALM gas according to a joint regression of metallicity against $M^*$ and $\Sigma^*$ (essentially, a global-local model), this study aims to detect the simple presence of ALM gas on a galaxy-by-galaxy basis in a galactocentric annulus between 1.25 \& 1.75 $R_e$ according to an elevated dispersion of the metallicity distribution function within that annulus, along with a steep radial metallicity decrement relative to the central value. We are unable to make direct comparisons between this work's most likely inflow hosts and the most favored candidates from \citet{hwang_2019_manga_almrs}, since a list of galaxies is not publicly available. Generally, though, both this study and \citet{hwang_2019_manga_almrs} note more pervasive inflow signatures at low galaxy stellar mass. This is in spite of varying methodologies for locating ALM/inflow-diluted gas; different strong-line metallicity indicators; different sample-level \& data-quality cuts; and different treatment of spaxel metallicity uncertainty. Indeed, it seems that low-mass galaxies are promising hosts for large inflows.

The interpretation of HI-rich galaxies as accretion hosts is not universal: using a sample of HI-extreme galaxies, whose neutral gas content has been calibrated to peer galaxies' optical $r$-band luminosity, \citet{lutz_20_HI-eXtreme-iii} concludes that extremely HI-rich galaxies have their overall enrichment history decelerated by their relatively high angular momentum. However, this conclusion is based on a \textit{lack of correlation} between radial metallicity gradients and HI content, contrary to this work's principal results (though that study tabulates gradients in physical coordinates, ${\rm dex ~ kpc}^{-1}$). Further study of the HI dynamics of MaNGA galaxies relative to their spatially-resolved chemical content could guide this disagreement towards a resolution.

The concentration of plausible inflow hosts at low stellar mass constitutes an opportunity for further investigation. We suggest targeting for high-resolution HI follow-up galaxies with wide metallicity distribution functions at fixed radius, pronounced metallicity gradients (or decrements), and large HI mass fractions. A table is provided to aid in choosing targets for HI follow-up: a full version is included in machine-readable format, and we show a sample in Table \ref{tab:galaxies}. Resolved radio observations targeted according to chemistry may reveal coincident low-metallicity, star-formation-driven line emission \& cold gas enhancement, indicating active accretion from an external gas reservoir.

%% file: acknowledgements.tex
\section*{Acknowledgements}
ZJP, CT, and ALS acknowledge NSF CAREER Award AST-1554877. This research made use of \texttt{Astropy}, a community-developed core \texttt{python} package for astronomy \citep{astropy}; and \texttt{matplotlib} \citep{matplotlib}, an open-source \texttt{python} plotting library.

The authors are grateful to the anonymous referee, whose feedback aided in improving the statistical rigor of this work and the overall quality of the manuscript.

Funding for the Sloan Digital Sky Survey IV has been provided by the Alfred P. Sloan Foundation, the U.S. Department of Energy Office of Science, and the Participating Institutions. SDSS acknowledges support and resources from the Center for High-Performance Computing at the University of Utah. The SDSS web site is www.sdss.org.

SDSS is managed by the Astrophysical Research Consortium for the Participating Institutions of the SDSS Collaboration including the Brazilian Participation Group, the Carnegie Institution for Science, Carnegie Mellon University, the Chilean Participation Group, the French Participation Group, Harvard-Smithsonian Center for Astrophysics, Instituto de Astrof\'{i}sica de Canarias, The Johns Hopkins University, Kavli Institute for the Physics and Mathematics of the Universe (IPMU) / University of Tokyo, the Korean Participation Group, Lawrence Berkeley National Laboratory, Leibniz Institut f\"{u}r Astrophysik Potsdam (AIP), Max-Planck-Institut f\"{u}r Astronomie (MPIA Heidelberg), Max-Planck-Institut f\"{u}r Astrophysik (MPA Garching), Max-Planck-Institut f\"{u}r Extraterrestrische Physik (MPE), National Astronomical Observatories of China, New Mexico State University, New York University, University of Notre Dame, Observat\'{o}rio Nacional / MCTI, The Ohio State University, Pennsylvania State University, Shanghai Astronomical Observatory, United Kingdom Participation Group, Universidad Nacional Aut\'{o}noma de M\'{e}xico, University of Arizona, University of Colorado Boulder, University of Oxford, University of Portsmouth, University of Utah, University of Virginia, University of Washington, University of Wisconsin, Vanderbilt University, and Yale University.